\documentclass{article}

\usepackage{arxiv}
\usepackage[utf8]{inputenc} 
\usepackage[T1]{fontenc}    
\usepackage{hyperref}       
\usepackage{url}            
\usepackage{booktabs}       
\usepackage{amsfonts}       
\usepackage{nicefrac}       
\usepackage{microtype}      
\usepackage{lipsum}		
\usepackage{graphicx}
\usepackage{natbib}
\usepackage{doi}
\usepackage{makecell}
\usepackage{multirow}
\usepackage{xcolor}
\usepackage{amsmath}

\begin{document}

\title{ADAM Challenge: Detecting Age-related Macular Degeneration from Fundus Images}

\date{} 					

\author{Huihui Fang, Fei Li, Huazhu Fu, Xu Sun, Xingxing Cao, Fengbin Lin, Jaemin Son, Sunho Kim, Gwenole Quellec, \\Sarah Matta, Sharath M Shankaranarayana, Yi-Ting Chen, Chuen-heng Wang, Nisarg A. Shah, Chia-Yen Lee, Chih-Chung Hsu, \\ Hai Xie, Baiying Lei, Ujjwal Baid, Shubham Innani, Kang Dang, Wenxiu Shi, Ravi Kamble, Nitin Singhal, Ching-Wei Wang, \\ Shih-Chang Lo, José Ignacio Orlando, Hrvoje Bogunović, Xiulan Zhang, Yanwu Xu, iChallenge-AMD study group\footnotemark[1]}

\renewcommand{\thefootnote}{\fnsymbol{footnote}}
\footnotetext[1]{H.~Fang and F.~Li contributed equally to this work. 

X.~Zhang and Y.~Xu are the corresponding authors (E-mail: zhangxl2@mail.sysu.edu.cn; xuyanwu@baidu.com). 

H.~Fang, F.~Li, H.~Fu, X.~Sun, X.~Cao, F.~Lin, X.~Zhang, and Y.~Xu co-organized the ADAM challenge. 

F.~Li, F.~Lin, and X.~Zhang are with State Key Laboratory of Ophthalmology, Zhongshan Ophthalmic Center, Sun Yat-sen University, Guangdong Provincial Key Laboratory of Ophthalmology and Visual Science,Guangzhou, China.

H.~Fang, X.~Sun, X.~Cao, and Y.~Xu are with Intelligent Healthcare Unit, Baidu Inc., Beijing, China.

H.~Fu is with the Institute of High Performance Computing, Agency for Science, Technology and Research, Singapore.

J.I.~Orlando is with CONICET, Yatiris lab of Pladema Institute, Tandil, Argentina.

H.~Bogunović is with Department of Ophthalmology and Optometry, Medical University of Vienna, Vienna, Austria.

J.~Son and S.~Kim are with VUNO Inc. Seoul, Republic of Korea.

G.~Quellec and S.~Matta are with Inserm, UMR 1101, Brest, F-29200 France. S.~Matta is also with Univ Bretagne Occidentale, Brest, F-29200 France.

S.M.~Shankaranarayana is with Zasti India Pvt. Ltd.

Y.-T.~Chen is with Graphen.Inc.

C.-H.~Wang is with Muen Biomedical and Optoelectronics Technologies Inc.

N.A.~Shah is with Department of Electrical Engineering, Indian Institute of Technology, Jodhpur, Rajasthan, India.

C.-Y.~Lee is with Department of Electrical Engineering, National United University, 
Taiwan

C.-C.~Hsu is with Institute of Data Science and Department of Statistics, National Cheng Kung University,
Taiwan.

H.~Xie and B.~Lei are with Shenzhen University, Shenzhen, China.

U.~Baid and S.~Innani are with SGGS Institute of Engineering and technology, Nanded, India.

K.~Dang and W.~Shi are with Suzhou Tisu Information Technology Co. Ltd.

R.~Kamble and N.~Singhal are with AIRA Matrix, Mumbai, India.

C.-W.~Wang and S.-C.~Lo are with Graduate Institute of Biomedical Engineering, National Taiwan University of Science and Technology, Taiwan.
}






\maketitle

\begin{abstract}
Age-related macular degeneration (AMD) is the leading cause of visual impairment among elderly in the world. Early detection of AMD is of great importance, as the vision loss caused by this disease is irreversible and permanent. Color fundus photography is the most cost-effective imaging modality to screen for retinal disorders. Cutting edge deep learning based algorithms have been recently developed for automatically detecting AMD from fundus images. However, there are still lack of a comprehensive annotated dataset and standard evaluation benchmarks. To deal with this issue, we set up the Automatic Detection challenge on Age-related Macular degeneration (ADAM), which was held as a satellite event of the ISBI 2020 conference. 
The ADAM challenge consisted of four tasks which cover the main aspects of detecting and characterizing AMD from fundus images, including detection of AMD, detection and segmentation of optic disc, localization of fovea, and detection and segmentation of lesions. 
As part of the ADAM challenge, we have released a comprehensive dataset of 1200 fundus images with AMD diagnostic labels, pixel-wise segmentation masks for both optic disc and AMD-related lesions (drusen, exudates, hemorrhages and scars, among others), as well as the coordinates corresponding to the location of the macular fovea. A uniform evaluation framework has been built to make a fair comparison of different models using this dataset. During the ADAM challenge, 610 results were submitted for online evaluation, with 11 teams finally participating in the onsite challenge. This paper introduces the challenge, the dataset and the evaluation methods, as well as summarizes the participating methods and analyzes their results for each task. In particular, we observed that the ensembling strategy and the incorporation of clinical domain knowledge were the key to improve the performance of the deep learning models.
\end{abstract}

\keywords{AMD detection \and optic disc segmentation \and fovea localization \and lesion segmentation}

\section{Introduction}
\label{sec:introduction}
The macula, located in the posterior pole of the retina, is closely related to both fine and color vision. Once lesions appear in this region, people will suffer from vision decline, dark shadows, or dysmorphia. Age-related macular degeneration (AMD) is a degenerative disorder that affects the macular region, mainly occurring in people older than 45 years old~\citep{haines2006functional}. The etiology of AMD is not fully understood, although it has been observed that multiple factors such as genetics, chronic photo destruction effect, and nutritional disorders are linked to it~\citep{ambatiMechanismsAgeRelatedMacular2012, zajkac2015dry, rapalli2019nanotherapies}. AMD can be divided into early, intermediate and advanced stages according to clinical characteristics~\citep{ferris2013clinical}. Early and intermediate AMD are mainly manifested with the appearance of drusen and pigmentary abnormalities, and patients suffering from these lesions usually have normal or nearly normal vision. Advanced AMD, on the other hand, is associated with a much more severe vision loss. It is usually classified into two types: geographic atrophy (also known as dry AMD) and neovascular AMD (also known as wet AMD)~\citep{lim2012age}. In advanced dry AMD, choroidal retinal atrophies appear, leading to impaired central vision. Wet AMD, on the other hand, is characterized by active neovascularization under retinal pigment epithelium, subsequently causing exudates, hemorrhages, and scarring. This will eventually cause irreversible damage to the retinal photoreceptors and rapid vision loss if left untreated~\citep{maruko2007clinical}. An early diagnosis of AMD is crucial for timely treatment and prognosis. Fundus photography is a widely-available non-invasive examination, which turns it an essential tool for retinal disease screening. In the context of AMD, ophthalmologists observe the state of the macular region with fundus photographs to determine whether there are lesions such as drusen, exudates, hemorrhages, scars, geographic atrophies, neovascularizations, etc. However, qualitative analysis of fundus images is experience-dependent and time-consuming.

\begin{figure}[!t]
\centering
\includegraphics[width=0.8\linewidth]{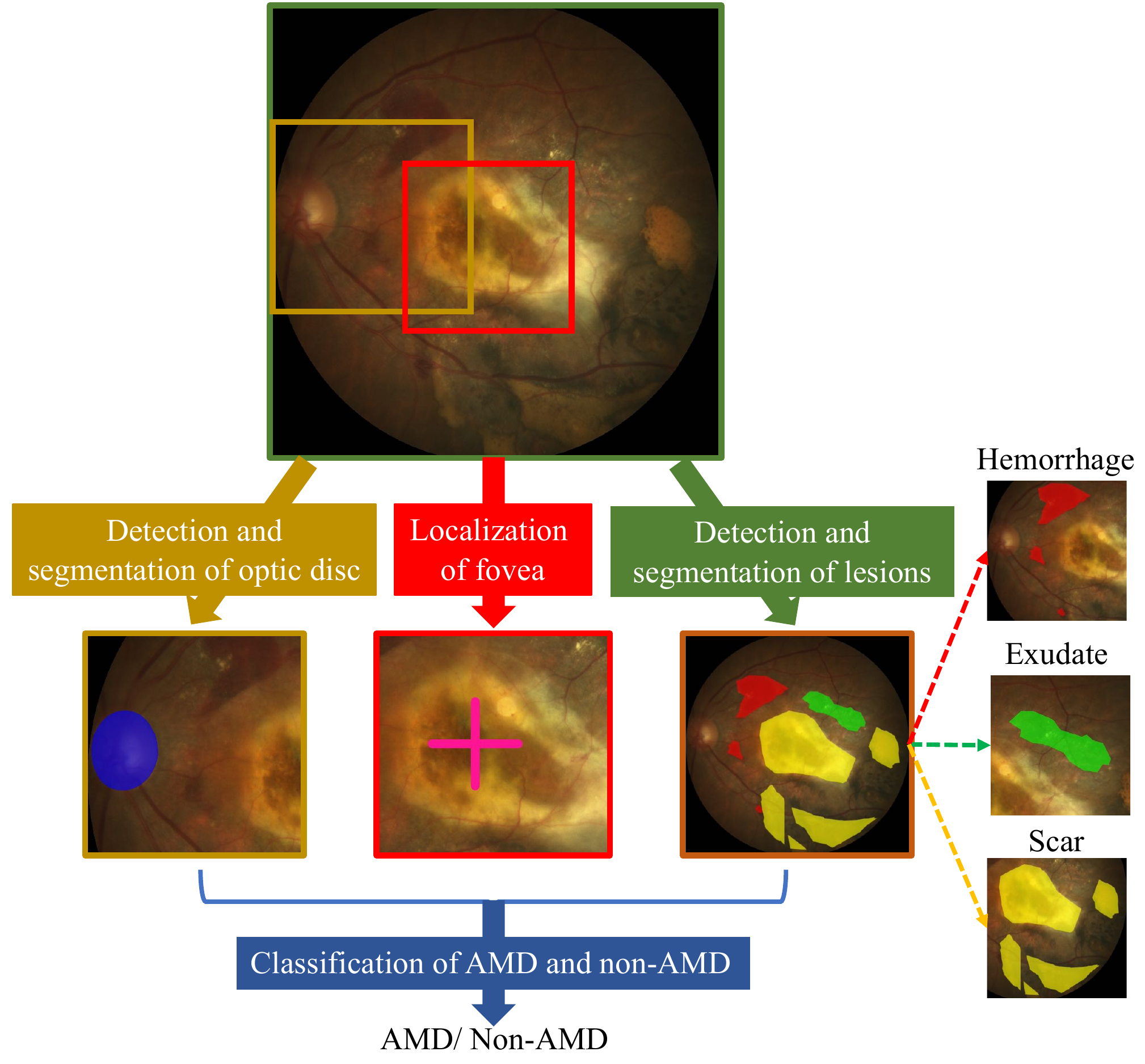}
\caption{The four tasks of the ADAM challenge. Classification of AMD and non-AMD, detection and segmentation of optic disc, localization of the fovea, and detection and segmentation of lesions. While there are five types of lesions that must be segmented in the challenge  (drusen, exudate, hemorrhage, scar, other), only three types are illustrated here.}
\label{fig1}
\end{figure}

With the rapid advancements and developments in image processing technologies, automated methods for fundus image analysis have been gradually introduced~\citep{fundus_survey}. Region growing or level sets algorithms for optic disc and cup segmentation~\citep{anitha2014region, thakur2019optic} or fovea localization methods based on hand-crafted features such as a concentric circular sectional symmetry measures~\citep{guoRobustFoveaLocalization2020} have proven accurate enough at this early stage to automatically, intuitively and quantitatively present the fundus structure to an expert. Other approaches focused on identifying disease related signs using similar techniques have produced methods for detecting drusen~\citep{akram2013automated} or hemorrhages~\citep{garcia2010assessment} among others. Disease detection models, on the other hand, were initially based on computing features from the entire image to subsequently categorize it using a dedicated classifier. Mookiah et al.~\citep{mookiah2014decision}, for instance, proposed an automated AMD detection system based on the discrete wavelet transform, in which first four-order statistical moments, energy, entropy, and Gini index-based features from the DWT coefficients were used as image-level descriptors. Altogether, these lesion segmentation and disease detection methods are able to provide a valuable AMD diagnostic message. However, their reliance on hand-crafted features designed to capture specific textural properties, grayscale intensities, or locations within the fundus structure in the image, had limited their overall performance and generalization ability. 
More recently, due to the cutting-edge developments in the field of deep learning, the methods of retinal structures analysis~\citep{Liu2019DeepAMD,wangPatchBasedOutputSpace2019,tabassumCDEDNetJointSegmentation2020,jiangJointRCNNRegionBasedConvolutional2020,xieEndtoEndFoveaLocalisation2020,maiyaRethinkingRetinalLandmark2020}, 
lesion extraction~\citep{vangrinsvenFastConvolutionalNeural2016,tanAutomatedSegmentationExudates2017,orlandoEnsembleDeepLearning2018,guoLSegEndtoendUnified2019,playoutNovelWeaklySupervised2019,engelbertsAutomaticSegmentationDrusen2019}, 
and disease prediction~\citep{burlinaAutomatedGradingAgeRelated2017,grassmannDeepLearningAlgorithm2018,Fu2018DiscAware,pengDeepSeeNetDeepLearning2019,He2020CABNet} based on fundus images have move towards more principled data-driven approaches. As a result, these methods can now take advantage of features automatically learned by deep neural networks, avoiding human biases in feature selection and/or design. However, supervised learning models require large amounts of annotated data to learn clinically relevant models. In natural image analysis tasks, pictures can be easily obtained using conventional cameras and then annotated by trained human cohorts. In medical image analysis, however, dedicated devices are needed to obtain the images, while the labels must be assigned by clinical experts, usually based not only on the image itself but on its associated records. This slows the development of novel tools to address AMD identification and characterization, unlike in other domains such as natural image classification, for which large scale public available datasets are available~\citep{russakovsky2015imagenet}. Furthermore, existing solutions are frequently presented and evaluated using their own protocols, rendering heterogeneous frameworks that cannot be easily compared to one another.

To address the above issues, we introduced the first \textbf{Automatic Detection challenge on Age-related Macular degeneration (ADAM)}. This competition\footnote{ \url{https://amd.grand-challenge.org}} is built on top of the successes of the REFUGE~\citep{orlandoREFUGEChallengeUnified2020a} and AGE~\citep{fuAGEChallengeAngle2020a} challenges, and was held in conjunction with the International Symposium on Biomedical Imaging (ISBI), 2020. According to the Age-Related Eye Disease Study (AREDS) \citep{davis2005age}, diagnosing AMD involves not only detecting diseased-related lesions but also analyzing their size and location to determine the severity of the disease. AREDS proposed a grid and standard circles based on the size of the optic disc and the localization of the fovea, which are used in assessing the lesions' properties. Hence, we believe that the automatic identification of fundus structure (optic disc and fovea) is also important in the clinical application due to their roles in AMD characterization. Therefore, the clinical goal of our challenge is not only the detection of AMD itself, but also the recognition of the two fundus structures and the segmentation of lesions. Our challenge featured then four basic tasks that can assist AMD detection and characterization, namely 
classification of AMD and non-AMD, detection and segmentation of optic disc, localization of fovea, and detection and segmentation of lesions (Fig.~\ref{fig1}).

AMD detection was posed as a binary classification task into AMD or non-AMD categories
rather than grading early, intermediate and advanced stages. These three grades were grouped into a single AMD category, while subjects without any disease sign were grouped as non-AMD~\citep{burlina2017automated}.
To train and evaluate methods for solving this and the other three tasks, we released a large dataset of 1200 fundus images, each of them associated with the following expert made annotations: a label of AMD classification, a segmentation mask for the optic disc, a location coordinate of the macular fovea, and five segmentation masks of disease related lesions: drusen, exudates, hemorrhages, scars, and others. In addition, to standardize a fair comparison of the results obtained by different methods, a unified evaluation framework was provided. 
In this paper, we introduce the ADAM challenge and the released dataset in detail, summarizing the methodologies proposed by the participating teams, and reporting their performances. Moreover, we discuss the impact on the performance of the deep learning based models when using ensemble methods or when incorporating additional datasets and/or prior medical knowledge. Finally, we discuss the clinical significance of these results in the context of AMD screening from fundus images.

\section{The ADAM challenge}
\label{sect:adam}
The ADAM challenge focuses on the investigation and development of algorithms associated with the diagnosis of AMD and the analysis of fundus photographs. Hence, we released a large dataset of 1200 annotated retinal fundus images. In addition, to enable a fair comparison of the results obtained by different algorithms, we designed a common evaluation framework. The challenge consisted of a preliminary (online) stage and a final (onsite) stage. During the preliminary stage, we released a training set and an online validation set for the model development and evaluation, respectively. There were 610 results submitted to the online evaluation platform, and 11 teams were invited to the final competition. In the final stage, an onsite test set was released to test the models.

\begin{table*}[!t] 
\small
\setlength\tabcolsep{3pt}
\centering
\caption{Summary of the ADAM challenge dataset labels and available annotations.}
\begin{tabular}{ccccccccccc}
\hline
\makecell[c]{Set} &
\makecell[c]{Num.\\samples} &
\makecell[c]{AMD/ \\Non-AMD} & 
\makecell[c]{Early/Inter-\\mediate/Late-dry/\\Late-wet AMD} &
\makecell[c]{With/o \\ Optic Disc} & 
\makecell[c]{With/o \\ Fovea} & 
\makecell[c]{With/o \\Drusen} & 
\makecell[c]{With/o \\Exudate} & 
\makecell[c]{With/o \\Hemorrhage} & 
\makecell[c]{With/o \\Scar} & 
\makecell[c]{With/o \\Other Lesions} \\ 
\hline
	Training      & 400 & 89/311    & 28/7/4/50              & 270/130                    & 396/4                 & 61/339                  & 38/362                   & 19/381                      & 13/387                & 17/383                                   \\ 
	Online & 400 & 89/311  & 18/7/3/61                & 265/135                    & 400/0                 & 49/351                  & 53/347                   & 30/370                      & 34/366                & 10/390                                   \\ 
	Onsite   & 400    & 89/311  & 24/9/3/53                & 286/114                    & 394/6                 & 44/356                  & 39/361                   & 23/377                      & 21/379                & 2/398                                    \\  
	\hline
	Total  & 1200    & 267/933     & 70/23/10/164            & 821/379                    & 1190/10               & 154/1046                & 130/1070                 & 72/1128                     & 68/1132               & 29/1171                                  \\ 
	\hline
\end{tabular}
\label{tab1} 
\end{table*}

\subsection{ADAM dataset}
\label{subsect:adamdata}

\begin{figure}[!t]
\centering
\includegraphics[width=1\linewidth]{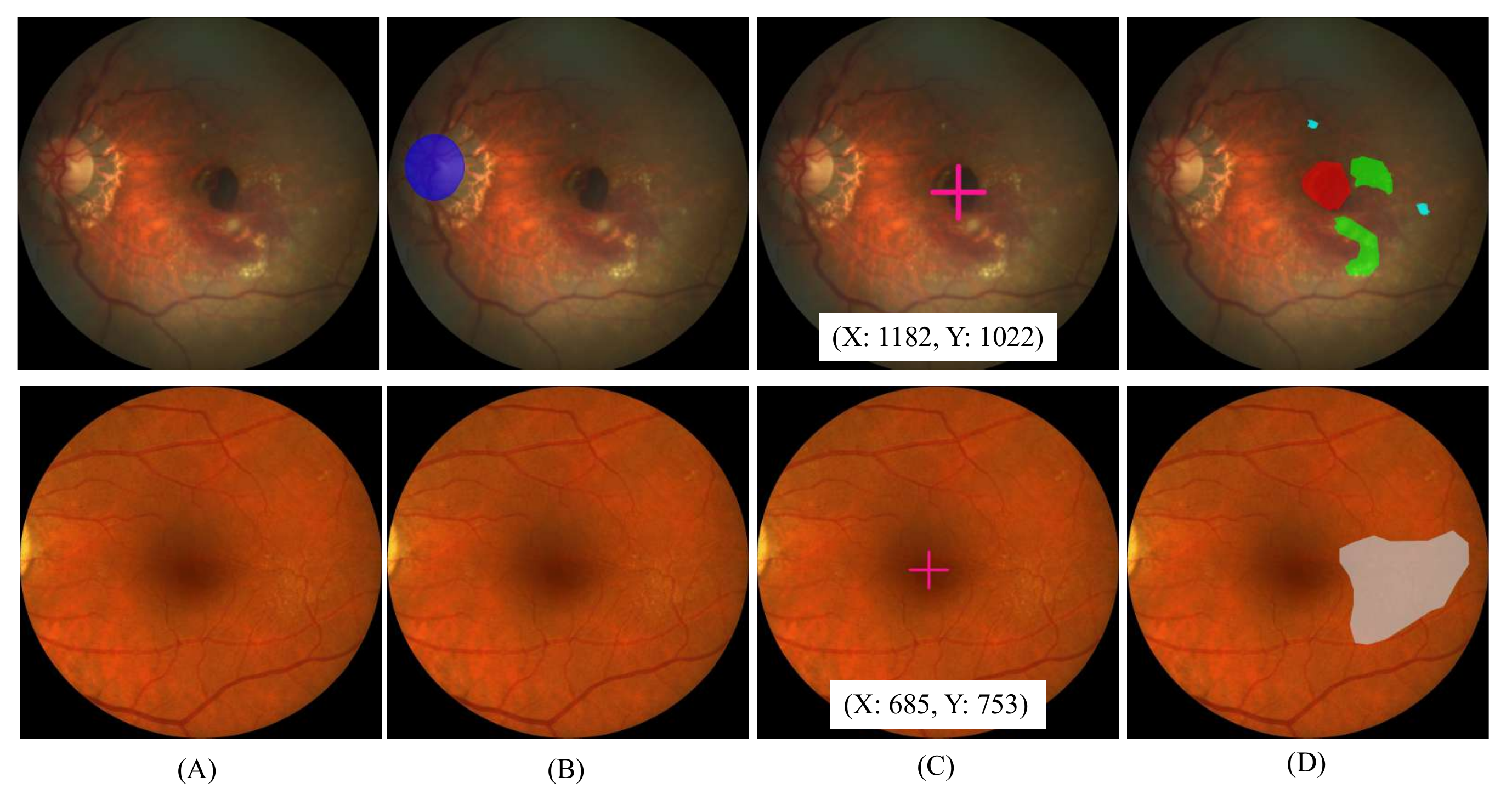}
\caption{Reference standards for all tasks in the ADAM challenge. Top row: AMD case. Bottom row: non-AMD case. From left to right: (A) original fundus image; (B) optic disc segmentation mask (blue) (notice that the optic disc is not visible in the non-AMD case); (C) fovea location coordinate annotation (pink cross); (D) lesion segmentation masks (light blue: drusen; green: exudate; red: hemorrhage; gray: others; no scars visible in the two samples).}
\label{fig2}
\end{figure}

The ADAM dataset consists of 1200 retinal fundus images stored in JPEG format with 8 bits per color channel, which are provided by Zhongshan Ophthalmic Center, Sun Yat-sen University, China. The fundus images were acquired by using a Zeiss Visucam 500 fundus camera with a resolution of 2124$\times$2056 pixels (824 images) and a Canon CR-2 device with a resolution of 1444$\times$1444 pixels (376 images). 
The collection process strictly followed the operating procedures of the fundus cameras. 
The examinations were made in a standardized darkroom and with the patients sitting upright. The center of the field of view of the photographs were either placed in the optic disc, the macula, or the midpoint of the optic disc and macula. All examinations were performed on Chinese patients (47\% female), who are at the ages of $53.19 \pm 15.59$ years old, in Zhongshan Ophthalmic Center. The final dataset was constructed by manually selecting  high-quality images from either the left or the right eye. The personal information of every image was removed for privacy. The resulting dataset was then divided into three parts: a training set (400 images), an online set (400 images), and an onsite set (400 images). Table \ref{tab1} summarizes the characteristics of the dataset. In addition to the binary category labels of AMD/not-AMD, we also provide pixel-wise segmentation masks of the full optic disc and lesions, as well as the location coordinates of the macular fovea (Fig.~\ref{fig2}). To the best of our knowledge, ADAM is the first dataset containing such a comprehensive set of labels for AMD analysis. 

The reference standard for AMD classification was obtained from clinical diagnosis results, based on both text information registered on the associated medical records--such as medical history and physical examination results--and imaging information, including fundus images and optical coherence tomography (OCT). Images labeled as AMD in our dataset correspond to any of early, intermediate, or advanced AMD cases~\citep{ferris2013clinical}, while the non-AMD category corresponds to samples without AMD, but that may have other retinal disorders.
In the training, online, and onsite sets, the AMD samples account for 22.25\% (89 images each). The sizes of early/intermediate/advanced-dry/advanced-wet AMD samples in these three datasets are 28/7/4/50, 18/7/3/61, and 24/9/3/53, respectively (see Table~\ref{tab1}). It is worth noting that the proportion of AMD samples in our dataset, especially corresponding to wet AMD, is higher than that in the real-world, which is mainly due to retrospectively retrieving scans from a hospital to which patients are referred with serious ocular disease. In addition, we also deliberately increased the number of the AMD category to partially reduce the effect of uneven class distribution in model training. Other retinal diseases, such as diabetic retinopathy, myopia and glaucoma, can be found in certain AMD and non-AMD samples. Initial manual pixel-wise annotations of the full optic disc and five lesions (drusen, exudate, hemorrhage, scar, and an extra category for `other lesions') were provided by 7 independent ophthalmologists who carefully reviewed and delineated the targets in all images. None of them had access to any patient information or knowledge of disease prevalence in the data to avoid any potential bias. The majority voting of the 7 annotations was taken to create a single disc or lesion segmentation mask per image, followed by a quality check by a senior specialist. When errors in the annotations were observed, this additional specialist analyzed each of the 7 segmentations, removed those that were considered erroneous in his/her opinion, and repeated the majority voting process with the remaining ones. 
Similarly, the initial coordinates of the fovea were obtained by 7 independent experts and the final coordinate was determined by averaging of the 7 annotations, with a senior specialist performing a quality check afterwards. When errors in the final coordinates were observed, the same correction method as performed with the segmentation masks was followed. The inter-rater agreement was considered in the dataset preparation. For the segmentation annotations, we referred to the work of Visser et al.~\citep{visser2019inter}, and adopted generalized conformity index (GCI)~\citep{kouwenhoven2009measuring} to measure the inter-rater agreement. The GCI of the 7 annotation results was 0.715, so their agreement was considered excellent according to the boundary value used in Visser et al.’s paper (GCI value of 0.7-1.0 are regarded as excellent~\citep{bartko1991measurement,zijdenbos1994morphometric}). For the fovea coordinates annotations which were continuous variables, we utilized intra-class correlation coefficient (ICC)~\citep{mcgraw1996forming} to measure the agreement according to Ranganathan’s paper~\citep{ranganathan2017common}. Note that in our study, the difference calculation of different values is changed to Euclidean distance calculation of different points when using ICC. Finally, the ICC of the fovea coordinates labeled by 7 annotators was 0.87993. According to the latest ICC boundary value~\citep{koo2016guideline} (ICC value of 0.75-0.9 are considered as good), the inter-rater agreement was good.

\subsection{Challenge Evaluation}
\label{challenge_evaluation}
Specific evaluation metrics and performance ranking rules were designed for every task in the ADAM challenge. After the teams submitted their predicted results, a unified evaluation method was used to calculate the effects of different models, in order to ensure a fair comparison. Since the evaluation metrics corresponding to individual tasks differ in scale and units, we did not directly calculate the ranking by averaging single tasks measurements. Instead, we first ranked the teams for each of the tasks, and then weighed the task ranks to obtain the overall ranking for the challenge. This process was previously followed in~\citep{orlandoREFUGEChallengeUnified2020a} to avoid bias in the final ranking toward any particular single task.

\subsubsection{Task 1: Classification of AMD and non-AMD images}
\label{subsubseq:classification}

Participants provided the estimated probability/risk of the image belonging to a patient diagnosed with AMD (value from 0.0 to 1.0). The classification results were compared to the clinical diagnosis of AMD. A receiver operating characteristic (ROC) curve was created across all the test set images and the area under the curve (AUC) was calculated. Each team received a rank (1=best) based on the obtained AUC value. 

\subsubsection{Task 2: Optic disc detection and segmentation}
\label{subsubseq:optic-disc}

Participants provided the segmentation results as one mask per testing image, with the segmented pixels labeled in the same way as in the reference standard (PNG files with 0: optic disc, 255: elsewhere). If the optic disc was detected to be not fully present, the pixel-wise labels in the segmentation were all set to be 255. The segmentation metrics were first calculated in every sample with optic disc, and then the results were averaged. The optic disc detection metric, on the other hand, was directly calculated among all test samples.
As previously done in other eye challenges~\citep{PALM, orlandoREFUGEChallengeUnified2020a}, the Dice coefficient was used to rank the teams for the segmentation task: 
\begin{equation}
\text{Dice}=\frac{2 \cdot |A\cap B|}{|A|+|B|}=\frac{2 \cdot \text{TP}} {2\text{TP}+\text{FP}+\text{FN}}.
\end{equation}
where $|A|$ and $|B|$ represent the number of pixels of the prediction and ground truth and $|A \cap B|$ is the  number of pixels in the overlap between the prediction and ground truth. TP, FP, and FN correspond to the amount of true positive, false positive, and false negative pixels in each image, respectively. Each team received a rank $R^\text{seg}_\text{disc}$ (1=best) based on the mean value of Dice over the testing images.   

In addition, an optic disc was assumed to be detected in an image if its segmentation had pixels being labeled as the optic disc. Accordingly, the $F_{1}$ score was calculated as the detection evaluation metric: 
\begin{equation}
F_{1} = \frac{2 \cdot P \cdot R}{P+R} =\frac{2 \cdot \text{TP}'}{2\text{TP}'+\text{FP}'+\text{FN}'}.
\end{equation}
where $P$ and $R$ represent precision and recall of the detection results among the testing images. $\text{TP}'$, $\text{FP}'$ and $\text{FN}'$ represent the numbers of true positive, false positive, and false negative detections of the optic disc on an image, respectively. Notice that, although Dice and $F_{1}$ are mathematically equivalent, their meaning differs for segmentation and detection. In particular, the true positives, false positives and false negatives in Dice for segmentation are computed at pixel-level, whereas those in $F_{1}$ score for detection are computed at image-level. Each team received a rank $R^\text{det}_\text{disc}$ (1=best) based on the obtained $F_{1}$ score. Because segmentation results could provide more relevant information for clinical applications (e.g. the optic disc size is used to determine the image grid), we set a higher weight for $R^\text{seg}_\text{disc}$. Thus, the final ranking for the optic disc detection and segmentation task was determined as follows: 
\begin{equation}
R_{disc} = 0.4 \cdot R^\text{det}_\text{disc} + 0.6 \cdot R^\text{seg}_\text{disc}.
\end{equation}

\subsubsection{Task 3: Fovea localization}
\label{subsubseq:fovea}

Participants submitted their localization results as $(X,Y)$ coordinates. If the fovea was invisible in the given image, both $X$ and $Y$ coordinates were set to 0. These values were compared to the reference standards, for all the images in the test set. The average Euclidean distance between the estimations and the ground truth was used to score each team (the lower, the better). A rank was then assigned to each participating team (1=best) based on this measure.

\subsubsection{Task 4: Lesion detection and segmentation} 
\label{subsubseq:lesion}

Participants were asked to provide five lesion segmentation results per image, following the same principle used for task 2. Hence, one image per testing image was requested, as PNG files with 0: lesion and 255: elsewhere. As for task 2, the evaluation of segmentations was calculated in the samples with real lesions, while the evaluation of detections was calculated in all the test samples.
Dice was used here to evaluate each lesion segmentation result independently. Each team received a rank $R_{i}^\text{seg}$ (1=best) for each lesion $i$ based on the mean value of the Dice measure over the testing images where the corresponding lesion was actually present. In addition, a specific lesion was assumed to be detected in an image if the submitted segmentation mask had pixels being labeled as that lesion. Accordingly, the $F_{1}$ score was calculated as a detection evaluation metric. Each team received a rank $R_{i}^\text{det}$ (1=best) for each lesion $i$ based on their detection evaluation measures. The evaluation score was then determined by weighting the two individual ranks obtained for each lesion. Hence, the evaluation ranking for lesion segmentation was determined by doing $R_{i}=0.4 \cdot R_{i}^\text{det} + 0.6 \cdot R_{i}^\text{seg}$, where $i$ represents drusen, exudate, hemorrhage, scar or other. The final ranking was finally computed as: 
\begin{equation} 
R_\text{lesion} = \sum_{i} R_{i},  
\end{equation} 
and used to determine the lesion detection and segmentation leaderboard (the smallest $R_\text{lesion}$ value, the better).

\subsubsection{Final evaluation}
\label{subsubseq:final-eval}

The final ranking of the ADAM challenge was calculated by the following equation: 
\begin{equation}
R = 0.3 \cdot R_\text{AMD} + 0.1 \cdot R_\text{disc} + 0.1 \cdot R_\text{fovea} + 0.5 \cdot R_\text{lesion},
\end{equation}where $R_\text{AMD}$, $R_\text{disc}$, $R_\text{fovea}$ and $R_\text{lesion}$ represent the ranks of the aforementioned four leaderboards. As the information of lesions is the most clinically relevant evidence for diagnosis, we gave the highest weight to $R_\text{lesion}$. AMD classification rank was also assigned with a high weight as it provides a direct diagnostic recommendation. Finally, the ranks of the remaining two tasks for fundus structure analysis were assigned the lowest weights. $R$ was used to determine the final leaderboard (1=best). In case of a tie, the rank of the classification leaderboard ($R_\text{AMD}$) had the preference. 

After the online evaluation, 11 teams attended the final onsite challenge during ISBI 2020. The onsite dataset was released for these teams and the final results were asked to be submitted within a 6 hours timeframe. The final ranking $R_\text{final}$ was then used to choose the best team, considering both online and onsite evaluation rankings: 
\begin{equation}
R_\text{final} = 0.3 \cdot R_\text{online} + 0.7 \cdot R_\text{onsite}.
\end{equation}As the teams were able to see the performance on the online dataset through the leaderboard during the preliminary stage of the challenge, they were able to adjust their models to work better on this dataset. Therefore, a higher weight was assigned to the onsite ranking to prize the generalization ability of the competing models~\citep{orlandoREFUGEChallengeUnified2020a,fuAGEChallengeAngle2020a}.

\section{Results}
\label{results}
\begin{table*}[t]
\small
\caption{Summary of the methods applied by the participating teams on the AMD classification task.}
\centering
\begin{tabular}{cccccc}
\hline
\makecell[c]{Rank}&   
\makecell[c]{Team} &    
\makecell[c]{Additional Dataset} &    \makecell[c]{Architecture} &    \makecell[c]{Ensemble} &    
\makecell[c]{Loss} \\
\hline
1&  VUNO EYE TEAM&  -&    EfficientNet&    \makecell*[c]{Self-ensemble, Concatenate 15 \\ finding feature maps} &    CE\\

2&    ForbiddenFruit&    ODIR&    \makecell*[c]{EfficientNet, DenseNet}&  \makecell*[c]{Ensemble 5 models with\\ designed formation} &    CE\\

3&    Zasti\_AI&    -&    \makecell*[c]{EfficientNet, Inception-ResNet,\\ ResNeXt, SENet} &    \makecell*[c]{Ensemble 8 network predictions\\ using averaging of posterior\\ probabilities} &    CE\\

4&
Muenai\_Tim&
-&
EfficientNet&
\makecell*[c]{Self-ensemble 3 local\\  minimal loss models \\ using averaging method} &
CE\\

5&
ADAM TEAM&
-&
\makecell*[c]{Xception,
Inception-v3,\\
ResNet50,
DenseNet101} &
\makecell*[c]{Ensemble 3 models,
using \\ averaging method} &
CE\\

6&
WWW&
-&
EfficientNet&
-&
Weighted CE\\

7 &
XxlzT&  
- &
\makecell*[c]{Autoencoder with ResNet50\\ as backbone} &
-&
CE\\

8&
TeamTiger&
-&
ResNet-101&
-&
CE   
\\     
9&
Airamatrix&
ODIR&
EfficientNet-B4&
-&
CE   
\\ 
\hline
\end{tabular}
\label{tab2} 
\end{table*}

\begin{table}[t]
\small
\caption{Evaluation of the AMD classification task on the onsite set.}
\centering{}
\begin{tabular}{c|c|c}
	\hline
	     Team      &  AUC   & Rank \\ \hline
	VUNO EYE TEAM  & 0.9714 &  1   \\ 
	
	ForbiddenFruit & 0.9592 &  2   \\ 
	  Zasti\_AI    & 0.9581 &  3   \\ 
	 Muenai\_Tim   & 0.9399 &  4   \\ 
	  ADAM-TEAM    & 0.9287 &  5   \\ 
	     WWW       & 0.9178 &  6   \\ 
	    XxlzT      & 0.9097 &  7   \\ 
	  TeamTiger    & 0.9086 &  8   \\ 
	  Airamatrix    & 0.8847 &  9   \\
	  \hline
\end{tabular}
\label{tab6} 
\end{table}

\begin{figure}[t]
\centering
\includegraphics[width=0.8\linewidth]{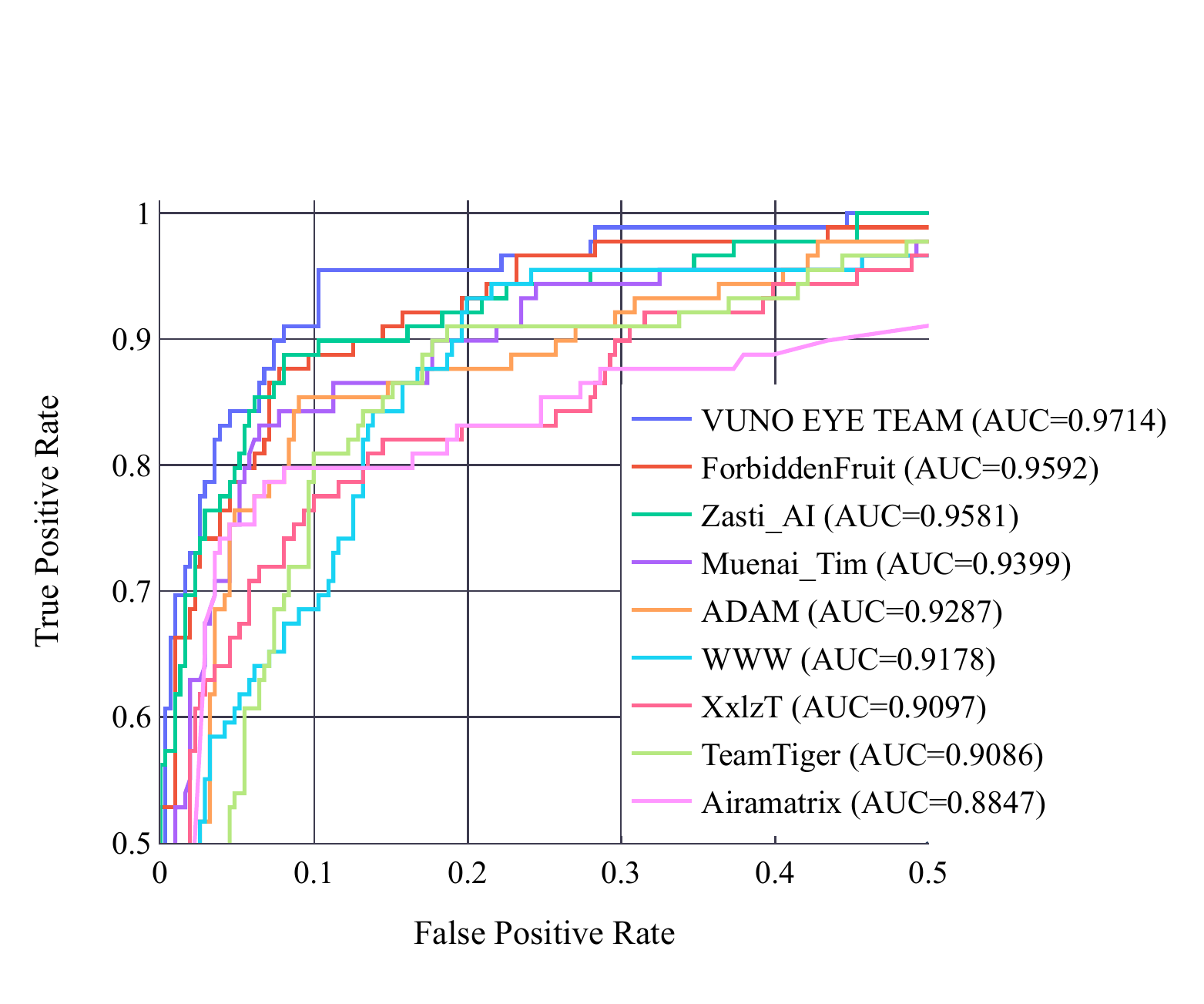}
\caption{ROC curves of the AMD classification results on the onsite challenge subset.}
\label{fig17}
\end{figure} 

We summarize in the sequel the methods and results of the participating teams in the ADAM challenge. Although 11 teams participated in the final stage, it is worth pointing out that not all of them took part in all the tasks. In particular, Voxelcloud team only participated in task 3. CHING WEI WANG (NUST) was not participating in task 1. 
The detailed leaderboards can be accessed on the ADAM challenge website at \url{https://amd.grand-challenge.org/ADAM-Finals/}.

\subsection{Classification of AMD and non-AMD}

As explained in Section~\ref{subsubseq:classification}, the purpose of this task is to determine whether the image corresponds to an AMD case or not. A brief summary of the methods adopted by the participating teams for this task is shown in Table~\ref{tab2}. A detailed description of each of them is available in the Appendix~\ref{appendix_A}. Notice that 5 out of 9 contributions applied ensembling strategies~\citep{zhouEnsemblingNeuralNetworks2002}. Ensemble methods are the most common approaches used by winners in Computer Vision competitions~\citep{nguyenFacialExpressionRecognition2019,khenedFullyConvolutionalMultiscale2019,sekiguchiEnsembleLearningHuman2020}. This approach is based on training multiple different architectures or the same neural network architecture under different settings, and combine their outputs to produce a single prediction per image. In general, this allows to improve accuracy with respect to the single-model counterparts.

\begin{table*}[ht]
\small
\caption{Summary of the methods applied by the participating teams on the optic disc detection and segmentation task.}
\centering
\begin{tabular}{m{0.5cm}<{\centering}m{1.5cm}<{\centering}m{2.7cm}<{\centering}m{4cm}<{\centering}m{2.5cm}<{\centering}m{2.5cm}<{\centering}m{1.5cm}<{\centering}}
\hline
\makecell[c]{Rank} &
\makecell[c]{Team} &
\makecell[c]{Additional Dataset} &
\makecell[c]{Architecture}&
\makecell[c]{Ensemble}&
\makecell[c]{Strategy}&
\makecell[c]{Loss} \\               
\hline
1&Airamatrix&-&\makecell[c]{FCN with ResNet50\\ as encoder}&-&\makecell[c]{Multi-task: jointly \\training with fovea \\segmentation}&CE\\

2&XxlzT&REFUGE, IDRiD&\makecell[c]{ResNet for classification,\\ U-Net for coarse segmentation,\\
DeepLab-v3+ for fine \\segmentation}& -& Multi-stage&
CE\\    

3&Forbidden Fruit&-&\makecell[c]{FPN with EfficientNet\\-B0 and -B2 as encoder}& \makecell[c]{Ensemble of 2 models \\ by averaging}&-& Focal, Dice\\    

4&WWW&RIGA, IDRiD&\makecell[c]{EfficientNet for classification,\\
U-Net for segmentation}&-&Multi-stage&CE\\

5&VUNO EYE TEAM&RIGA, IDRiD, REFUGE, PALM&\makecell[c]{U-Net with EfficientNet \\ as encoder}&Self-ensemble of 5 models at different epochs with majority voting &Multi-task: jointly training with OD segmentation based on vessel mask&
CE\\

6&TeamTiger&-&\makecell[c]{U-Net with EfficientNet B7\\ as encoder}&-&-&Jaccard\\

7 &ADAM-TEAM&-&\makecell*[c]{U-Net with Inception-v3,\\ EfficientNet-B3, ResNet50,\\ DenseNet101 as encoder}&\makecell*[c]{Ensemble of 4 \\models using \\averaging}&-&BCE, Dice\\    

8&Zasti\_AI&REFUGE&U-Net with ResNet as encoder&
-&-&CE\\

9&Muenai\_Tim&-&EfficientNet-B0 for classification, U-Net++ and EfficientNet-B7 for segmentation&Ensemble of 2 models for segmentation
&Multi-stage&BCE, Dice\\

10 & \makecell*[c]{CHING WEI WANG\\(NTUST)}&-&\makecell[c]{FCN with VGG16 \\as encoder}&-&-&BCE\\

\hline
\end{tabular}
\label{tab3} 
\end{table*}

\begin{table}[t]
\small
\centering
\caption{Evaluation of the optic disc detection and segmentation results, in terms of Dice (segmentation) and $F_{1}$ (detection), and final ranking on the task.}
\begin{tabular}{c|c|c|c}
	\hline
	     Team      &  Dice  & $F_{1}$ & Rank \\ \hline
	    XxlzT      & 0.9486 & 0.9913  &  1   \\ 
	    
	  Airamatrix   & 0.9475 & 0.9862  &  2   \\ 
	ForbiddenFruit & 0.9420 & 0.9912  &  3   \\ 
	     WWW       & 0.9445 & 0.9793  &  4   \\ 
	  TeamTiger    & 0.9429 & 0.9792  &  5   \\ 
	VUNO EYE TEAM  & 0.9370 & 0.9894  &  6   \\ 
	  ADAM-TEAM    & 0.9417 & 0.9675  &  7   \\ 
	 Zasti\_AI    & 0.9020 & 0.9843  &  8   \\ Muenai\_Tim    & 0.9294 & 0.9737  &  9   \\
	 CHING WEI WANG (NTUST) & 0.9224 & 0.9375 & 10 \\
	  \hline
\end{tabular}
\label{tab7} 
\end{table}

Table~\ref{tab6} summarizes the AUC values obtained by each team on this task, and their corresponding rank. All teams except Airamatrix reached an AUC over 0.9 on the onsite dataset, as shown in Fig.~\ref{fig17}. When the false positive rates (FPR) are 0.2, only the true positive rates (TPRs) of XxlzT, ADAM-TEAM, and Airamatrix teams were less than 0.9, indicating that the false positives of most teams could be kept relatively low while still meeting the clinical requirements.

\begin{figure}[t]
\centering
\includegraphics[width=1\linewidth]{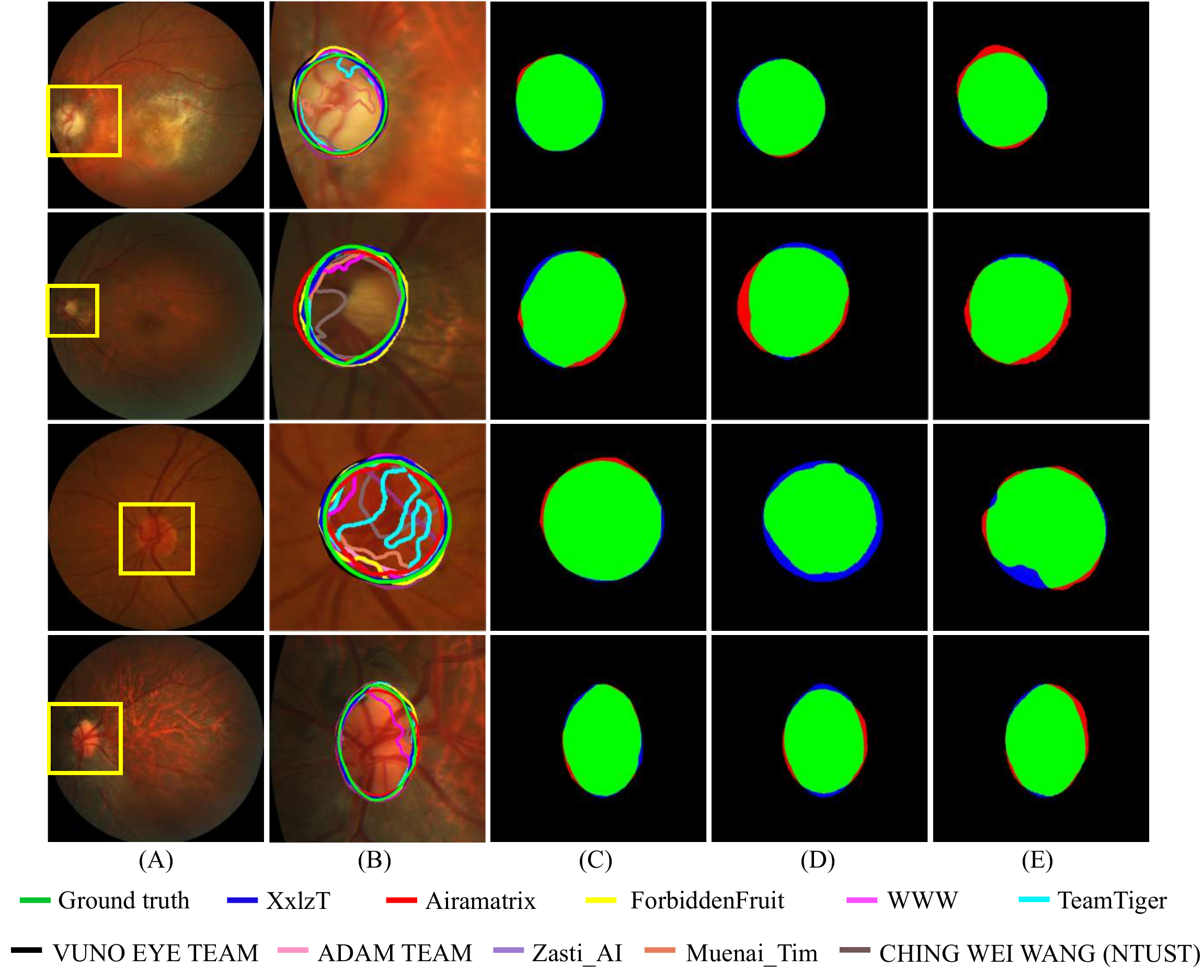}
\caption{Qualitative examples of the optic disc segmentation results. (A) Original fundus image, (B) Boundaries of the segmented optic disc obtained by the ten teams (color codes are provided in the legend), (C-E) Segmentation results of the top three teams 
compared with the ground-truth (green: true positive, blue: false negative, red: false positive).}
\label{fig20}
\end{figure}  

\subsection{Optic disc detection and segmentation}
The purpose of this task is to detect whether there is a complete optic disc present in the image (detection task), and, if that is the case, to obtain a pixel-wise segmentation of it (segmentation task) (Section \ref{subsubseq:optic-disc}).
The methods used by the participating teams are summarized in Table~\ref{tab3}. By reviewing their contributions, we divided them into three main categories: direct approaches that only segmented the optic disc, multi-step approaches which first detected and then segmented the optic disc, and multi-task learning models that were jointly trained to perform both tasks simultaneously. Further details can be found in the Appendix~\ref{appendix_B}. 
The evaluation results are shown in Table~\ref{tab7}. 
It can be seen that XxlzT obtained the best score in terms of Dice and $F_{1}$ (Dice$=0.9486$, $F_{1}=0.9913$). According to the ranking rules, the second and third-ranked teams are Airamatrix and Forbiddenfruit. Mann-Whitney U hypothesis tests with $\alpha=0.05,n_{1}=n_{2}=400$ were performed on these top three-ranked teams to compare the statistical significance of the differences in their results, using their obtained Dice values. We found that the XxlzT team significantly outperformed the Airamatrix ($p=3.52\times10^{-5}$) and Forbiddenfruit teams ($p=3.97\times10^{-7}$).

Fig.~\ref{fig20} includes qualitatively examples of the results for a subset of fundus pictures, including the boundaries of the predicted optic disc of all participating teams (B) and the segmentations of the top three teams (Airamatrix --C--, XxlzT --D--, and ForbiddenFruit --E--), compared with the ground truth. The first two rows show AMD images, and the last two rows show non-AMD cases. Regardless of AMD being present or not, all teams could approximately segment the optic disc in this sample. However, when the contrast of the optic disc is low due to illumination artifacts (row 3), the segmentations become more variable. In any case, results of the top three teams were relatively robust on this scenario.

\subsection{Fovea localization}
\begin{table*}[t]
\small
\caption{Summary of the methods applied by the participating teams on the fovea localization task.}
\centering
\begin{tabular}{m{0.5cm}<{\centering}m{1.5cm}<{\centering}m{2.7cm}<{\centering}m{3cm}<{\centering}m{3cm}<{\centering}m{3cm}<{\centering}m{1.5cm}<{\centering}}
\hline
Rank&
Team &
Additional Dataset &
Architecture&
Ensemble&
Strategy&
Loss \\               
\hline
1&
VUNO EYE TEAM&
IDRiD, REFUGE, PALM&
U-Net with EfficientNet as encoder&
Ensemble of 2 models by averaging&
Add vessel information, segmentation framework &
CE\\

2&
Forbidden Fruit&
-&
Segmentation: three FPNs with EfficientNet-B0, -B1 and -B2 as encoders; Regression: VGG-19, Inception-v3  and ResNet-v2-50&
For both segmentation and regression: ensemble of 3 models&
Segmentation and regression&
CE, MSE 
\\    

3&
Voxelcloud&
IDRiD, ARIA,
Proprietary dataset&
Nested U-Net&
Ensemble of 20 models by averaging&
Segmentation and regression; add vessel information&
Dice, L2
\\    

4&
Airamatrix&
-&
FCN-ResNet50&
-&
Joint training with OD segmentation&
CE
\\

5&
Zasti\_AI&
-&
GAN&
-&
Distance map generation&
MSE
\\

6&
WWW&
-&
Segmentation: Two U-Net, Mask-RCNN; Regression: ResNet&
Ensemble of different models by averaging&
Segmentation and regression&
CE, MSE
\\

7 &
Muenai\_Tim&
-&
EfficientNet-B0, -B7&
-&
Classification, regression&
CE, MAE
\\    
8 & \makecell*[c]{CHING WEI \\WANG(NTUST)} & - & \makecell[c]{FCN with VGG16 \\as encoder} & - & Segmentation & BCE
\\

9&
TeamTiger&
-&
EfficientNet-B7&
-&
Regression&
MSE
\\ 

10&
ADAM-TEAM&
-&
YOLO-v2&
-&
Using the common object detection method&IOU, BCE,
MSE
\\ 

11&
XxlzT&
REFUGE, IDRiD&
Faster RCNN&
-&
Using the common object detection method&
BCE, Smooth L1
\\ 

\hline
\end{tabular}
\label{tab4} 
\end{table*}

\begin{table}[!t]
\small
\centering
\caption{Evaluation of the fovea localization task, in terms of Euclidean Distance (ED).}
\begin{tabular}{c|c|c}
	\hline
	Team            & ED       & Rank \\ \hline
	VUNO EYE TEAM   & 18.5538  & 1    \\ 
	
	ForbiddenFruit  & 19.7074  & 2    \\ 
	Voxelcloud Team & 25.2316  & 3    \\ 
	Airamatrix      & 26.1720  & 4    \\ 
	Zasti\_AI       & 28.8555  & 5    \\ 
	WWW             & 36.3596  & 6    \\ 
	Muenai\_Tim     & 69.0398  & 7    \\ 
	CHING WEI WANG(NTUST) & 109.0659 & 8\\
	TeamTiger       & 192.6720 & 9    \\
	ADAM-TEAM       & 205.7886 & 10    \\
	XxlzT       & 284.0335 & 11    \\
	\hline
\end{tabular}
\label{tab8} 
\end{table}

\begin{figure}[t]
\centering
\includegraphics[width=1\linewidth]{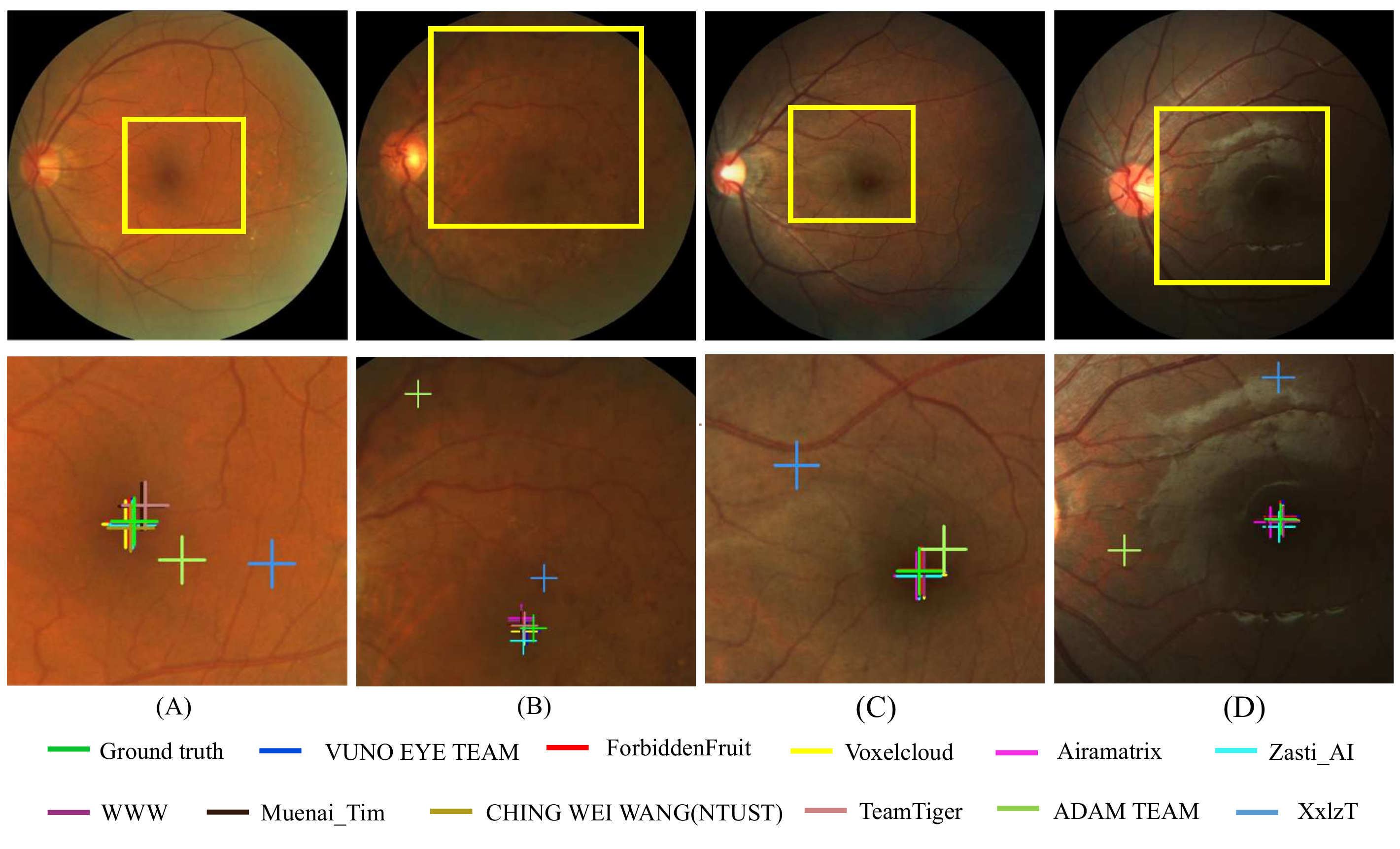}
\caption{Qualitative examples of the fovea localization results for all the participating teams. Yellow boxes indicated on the fundus images in the top row correspond to the enlarged areas illustrated in the bottom row. Predicted coordinates are shown with crosses in the respective team color.}
\label{fig22}
\end{figure}

The purpose of this task is to predict the coordinates of the fovea location (Section~\ref{subsubseq:fovea}). Table~\ref{tab4} summarizes the methods used by the participating teams on this task. Again, we can briefly classify these methods into regression-based, segmentation-based, and object detection-based models. Further details about these methods are provided in the Appendix~\ref{appendix_C}. The evaluation results in terms of the mean ED are summarized in Table \ref{tab8}.
The VUNO EYE TEAM achieved the best performance, with an ED of 18.5538 pixels. ForbiddenFruit and Voxelcloud teams achieved the second and third-best performances, with ED values of 19.7074 and 25.2316 pixels, respectively. The statistical significance of the differences in performance of the top three teams was studied using Mann-Whitney U hypothesis tests with $\alpha=0.05,n_{1}=n_{2}=400$. The differences in the ED values achieved by the VUNO EYE TEAM were statistically significant with respect to the Voxelcloud team ($p=8.82\times10^{-7}$), while not to the ForbiddenFruit team ($p=0.567$). Fig.~\ref{fig22} shows four examples of the fovea localization results of the participating teams. The samples (A) and (B) are AMD images, and the remaining samples are non-AMD cases. The 11 participating teams were able to get better fovea localization results when the macular region had a higher contrast and cleaner texture, regardless of AMD or non-AMD images.

\subsection{Lesion detection and segmentation}

\begin{table*}[!t]
\small
\caption{Summary of the methods applied by the participating teams on the lesion detection and segmentation task.}
\centering
\begin{tabular}{m{0.5cm}<{\centering}m{1.5cm}<{\centering}m{2.7cm}<{\centering}m{3.5cm}<{\centering}m{2.5cm}<{\centering}m{3cm}<{\centering}m{1.5cm}<{\centering}}
\hline
Rank&
Team &
Additional Dataset &
Architecture&
Ensemble&
Post-processing&
Loss \\               
\hline
1&
VUNO EYE TEAM&
-&
U-Net with EfficientNet as encoder&
Self-ensemble: concatenate 15 finding feature maps as encoder output&
-&
CE\\

2&
Zasti\_AI&
-&
U-Net with Residual blocks as encoder&
-&
Lesion area constraint&
CE
\\    

3&
WWW&
-&
DeepLab-v3 with ResNet&
Ensemble of 2 models&    
Region fill&
CE
\\    

4&
Airamatrix&
DiaretDB1&
DeepLab-v3 with Xception&
-&
-&
CE
\\

5&
Forbidden Fruit&
-&
FPN with EfficientNet&
Ensemble of 2 models with designed formation&
AMD score constraint&
Focal, Dice
\\

6&
Muenai\_Tim&
-&
Nest-U-Net, FPN, DeeplabV3 with EfficientNet-B7 and B3 as backbone&
Ensemble of different models using majority voting&
-&
CE, Dice
\\

7 & \makecell[c]{CHING WEI \\WANG(NTUST)} & - & \makecell[c]{FCN with VGG16\\ as encoder} & - & - & BCE
\\
8 &
ADAM-TEAM&
-&
U-Net with Inception-v3, EfficientNet-B3, ResNet50, DenseNet101 as encoder&
Ensemble of different models using averaging method&
Contour filling& CE
\\    

9&
TeamTiger&
-&
U-Net with EfficientNet-B0 as encoder&
-&
-&
Jaccard
\\

10&
XxlzT&
-&
ResNet50 for classification, DeepLab-v3+ for segmentation&
-&
-&
BCE
\\

\hline
\end{tabular}
\label{tab5} 
\end{table*}

\begin{table*}[t]
\small
    \centering
    \caption{Evaluation of the lesion detection ($F_{1}$) and segmentation (Dice) results obtained by each team for each type of lesion. The highest Dice and $F_{1}$ values are highlighted in bold.}
    \begin{tabular}{c|c|c|c|c|c|c|c|c|c|c|c}
    \hline
    \multirow{2}{*}{Team} & \multicolumn{2}{c|}{Drusen} & \multicolumn{2}{c|}{Exudate} & \multicolumn{2}{c|}{Hemorrhage} & \multicolumn{2}{c|}{Scar} & \multicolumn{2}{c|}{Other} & \multirow{2}*{Rank}\\ \cline{2-11}
    ~ &Dice&  $F_{1}$ &  Dice &$F_{1}$ &   Dice &$F_{1}$ &   Dice &$F_{1}$ &   Dice& $F_{1}$&~\\
    \hline
    VUNO EYE TEAM& 0.4838 & \textbf{0.6316} & 0.4154 & \textbf{0.5688} & \textbf{0.4303} & 0.7307&  0.4051 & 0.7027&  0.2852&  0.0714& 1
    \\               
    
    Zasti\_AI &   \textbf{0.5549}&  0.4972 & \textbf{0.4337} & 0.4965 & 0.2400 & 0.3614 & 0.5466 & 0.4598 & 0.2668 & 0.0290 & 2
    \\
    
    WWW &0.4836 & 0.4018 & 0.3174 & 0.5581 & 0.2190&  0.6000 & \textbf{0.5807}&  0.7273&  0.0344&  0.1176&  3
    \\
    
    Airamatrix & 0.3518&  0.5674 & 0.2606 & 0.4673 & 0.2257 & 0.2466 & 0.4080 & 0.6500 & \textbf{0.6906} & 0.1818 & 4
    \\
    
    ForbiddenFruit & 0.4007&  0.5443&  0.2866 & 0.5155 & 0.2079&  \textbf{0.8293}&  0.5639 & \textbf{0.8511} & 0.1224&  0.0085&  5
    \\        
    
    Muenai\_Tim & 0.4483 & 0.5535&  0.2140&  0.4634&  0.2164 & 0.6038 & 0.4248 & 0.6957&  0.1595&  0.0910&  6
    \\
    CHING WEI WANG(NTUST) & 0.2903 & 0.5153 & 0.2035 & 0.4724 & 0.1402 & 0.4932 & 0.2996 & 0.5806 & 0.2118 & \textbf{0.4724} & 7    
    
    \\

    ADAM-TEAM &   0.3260 & 0.1986 & 0.3256 & 0.1785 & 0.1815 & 0.1093 & 0.4038 & 0.1002 & 0.0698&  0.0100 & 8
    \\  
    
    TeamTiger &  0.3260 & 0.1982 & 0.3256 & 0.1777&  0.1815&  0.1087 & 0.4038&  0.0998&  0.0698&  0.0100&  9
    \\
    
    XxlzT & 0.0157 & 
    0.0556 &
    0.1593 &
    0.4096 &
    0.0186 &
    0.0690 &
    0.1014 &
    0.2963 &
    0.0000 &
    0.0000 &
    10 
    \\
    \hline
    \end{tabular}
\label{tab9} 
\end{table*}

\begin{figure}[!t]
\centering
\includegraphics[width=1\linewidth]{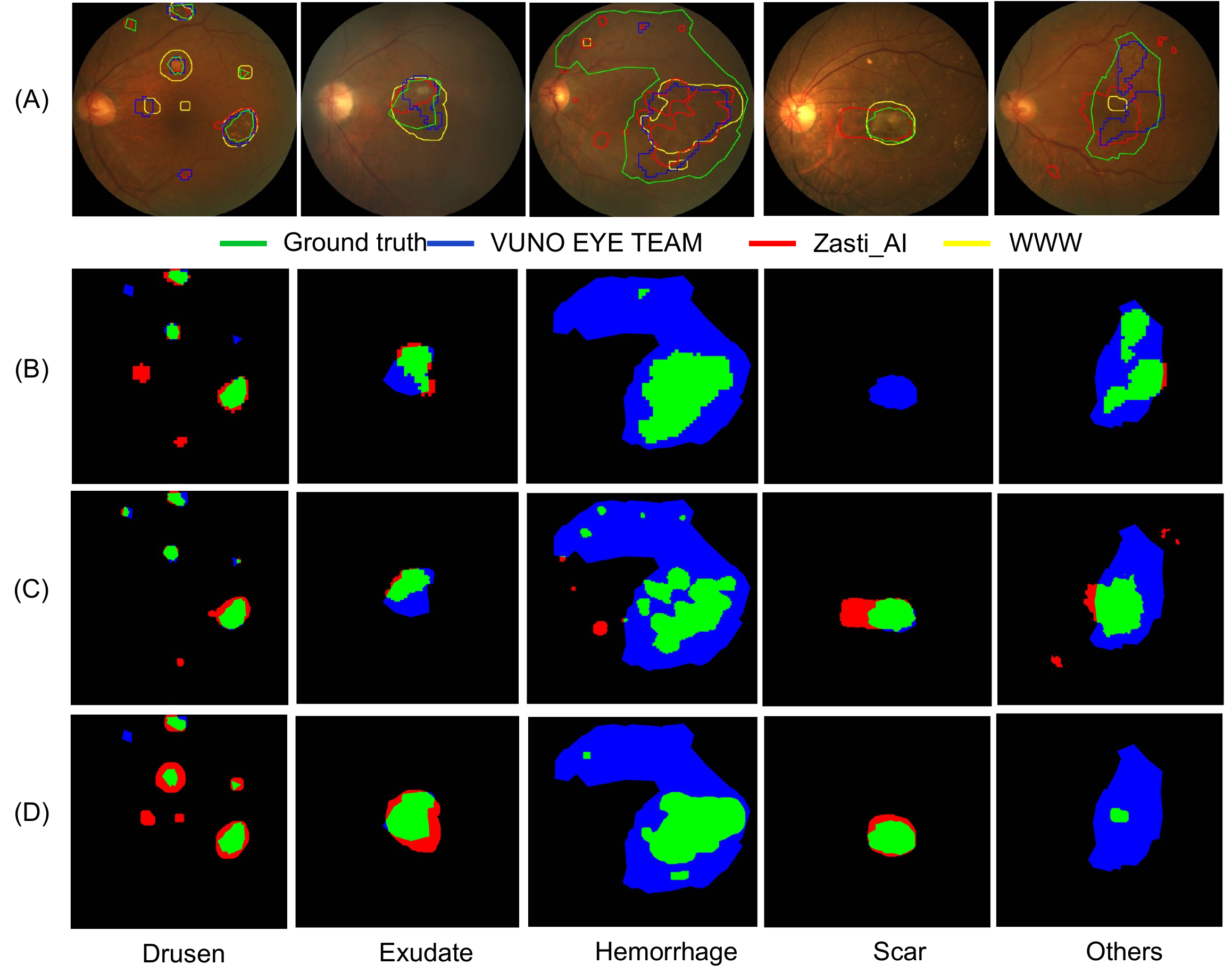}
\caption{Qualitative examples of lesion segmentation results obtained by the top three teams. From left to right: results for drusen, exudate, hemorrhage, scar and other lesions segmentations.  From top to bottom: (A) Original fundus images and boundaries corresponding to the segmented regions, (B-D) Segmentation masks of the top three teams, compared with the ground-truth (green: true positive, blue: false negative, red: false positive).}
\label{fig23}
\end{figure}

The objective of this task was to detect if the images contained drusen, exudate, hemorrhage, scars, or other lesions not listed, and to segment them if visible (Section~\ref{subsubseq:lesion}). A summary of the methods used by the participating teams is shown in Table~\ref{tab5}. Notice that U-Nets, DeepLab-v3, and FPN were the most commonly used architectures. The interested reader could refer to the Appendix~\ref{appendix_D} for further details. Table~\ref{tab9} provides the detailed evaluation in terms of Dice and $F_{1}$. VUNO EYE TEAM, Zasti\_AI, and WWW achieved the top three best performance according to our rank rule described in Section~\ref{subsubseq:lesion}. In particular, Zasti\_AI obtained the best results for the drusen and exudate segmentation tasks, with Dice values of 0.5549 and 0.4337, respectively. For the drusen and exudate detection, VUNO EYE TEAM achieved the best $F_{1}$ scores values of 0.6316 and 0.5688, respectively. This indicates that the method of VUNO EYE TEAM had advantages in detecting drusen and exudate presence, but that its pixel-wise segmentation results should be improved. For hemorrhage segmentation and detection, VUNO EYE TEAM and ForbiddenFruit had the best performance, with Dice of 0.4303 and $F_{1}$ of 0.8293, respectively. Finally, for detecting and segmenting scars and other non-listed lesions, the teams WWW and Airamatrix obtained the best segmentation results, and ForbiddenFruit and CHING WEI WANG(NTUST)  gained the best detection results, respectively.

Fig.~\ref{fig23} depicts some qualitative examples of segmentation results obtained by the three top ranked teams for each specific type of lesion. From the first column, it could be seen that the Zasti\_AI team produced the best result, while VUNO EYE TEAM and WWW oversegmented the drusen area. In the exudates example, on the other hand, WWW retrieved more true positive pixels than the other two teams, although at the cost of a higher number of false positives. Hemorrhages, which are widely distributed on the exemplary fundus image, are observed not to be fully captured by the winning methods. For the scar lesion segmentation, the result of the WWW team was the best one, which is consistent with the quantitative results in Table~\ref{tab9}. For segmenting lesions not listed in the other categories, we can see that all the top three teams produced several false negative pixels, especially the WWW team.

\section{Discussion}

In this section, we elaborate in detail the common properties of the proposed methods and their impact on the final results. In Section \ref{subsec:ensemble} we analyze the ensembling strategies proposed by the top-ranked teams. In Section \ref{subsec:class-imbalance} we study how much the variations in AMD stage and the amount of images with different types of lesions affected the outputs of the teams. The effect of incorporating additional training data or prior clinical knowledge are analyzed in Sections \ref{subsec:additional-datasets} and \ref{subsec:clinical-knowledge}, respectively. Finally, the clinical implications of the results achieved in the ADAM challenge are discussed in Section \ref{subsec:clinical-implications}.


\subsection{Model ensembling}
\label{subsec:ensemble}

As each neural network model has its own architectural characteristics and is trained under its own set of hyperparameters, it is expected for them to produce different predictions (and consequently, errors) when applied to the same set of inputs. This variability between models is exploited by ensembling methods, which utilize their complementarity to offset the variance by training multiple models and combining their predictions. As a result, predictions are less sensitive to the individual training settings and model characteristics, yielding usually more stable and accurate outputs~\citep{brownlee2018ensemble}. In general, there exists three ways to build variable models for ensembling: training a single architecture with different portions of data or datasets (adopted by VUNO EYE TEAM in tasks 1 and 4); training different models on the same datasets (as done by ForbiddenFruit team in task 1 and VUNO EYE TEAM in task 2); or simply using the same model and data but varying the training settings. Most of the participating teams chose the second method in the challenge as it does not require more datasets and allows obtaining more variable predictions. Once the outputs of all the constitutive models are collected, the ensemble strategies mainly included averaging and majority voting. Other specialized strategies can also be adopted, as was done by ForbiddenFruit team in task 1.

In the classification task of our challenge, the five top-ranked teams used ensemble methods, with the top three teams using responses from at least 5 models. It is important to note that the first team used a proprietary dataset of 15 lesion annotations and trained 15 models that had extracted lesion features associated with the AMD disease, which enabled better AMD classification results. For fovea localization, ensemble methods were also used for the top three teams, with the first team using 2, the second 6, and the third 20 models in total. The first and third methods were supplemented with additional vasculature information, while the second method only ensembled the outputs of individual regression networks. It can be inferred that increasing the number of ensembled methods might negatively affect the performance after reaching a certain, task-dependent level. Finding the appropriate number of models presents then a calibration experiment in itself, to be evaluated using a held-out portion of the data.

It must be pointed out, however, that the overall performance of the constitutive models for assembling must be adequate in order to ensure good final results. Some teams tailored these models to focus on specific details of the image, to further improve the overall response. In the optic disc segmentation task, for instance, the first team used the relevance of the positions of optic disc and macula in the fundus images to adjust the segmentation model. The second team, on the other hand, adopted a coarse-to-fine strategy to improve the final results. Although the third team used the ensembling method, their models dealt with the disc segmentation directly from the original image, hence the effect was not as strong as with the other two strategies.
\begin{table*}[t]
\setlength\tabcolsep{3pt}
  \centering
  \caption{Results obtained by ensembling the outputs of the top three ranked teams from each task. Numbers in green indicate a metric better than the winning method, while red values indicate that the results are less accurate.}
    \begin{tabular}{cccccccccccccc}
    \hline
    \multirow{2}[0]{*}{\makecell[c]{AMD \\Classification}} & \multicolumn{2}{c}{\multirow{2}[0]{*}{\makecell[c]{Disc Detection\\ and Segmentation}}} & \multirow{2}[0]{*}{\makecell[c]{Fovea \\Localization}} &
    
    \multicolumn{10}{c}{Lesion Detection and Segmentation} \\
    \cline{5-14}
          & \multicolumn{2}{c}{} &       & \multicolumn{2}{c}{Drusen} & \multicolumn{2}{c}{Exudate} & \multicolumn{2}{c}{Hemorrhage} & \multicolumn{2}{c}{Scar} & \multicolumn{2}{c}{Others} \\
          \hline
    AUC   & DICE  & $F_{1}$    & ED    & DICE  & $F_{1}$    & DICE  & $F_{1}$    & DICE  & $F_{1}$    & DICE  & $F_{1}$    & DICE  & $F_{1}$ \\
    \hline
    \textcolor{red}{0.9702} & \textcolor[rgb]{ 0,  .69,  .314}{0.9519} & \textcolor[rgb]{ 0,  .69,  .314}{0.9930} & \textcolor[rgb]{ 0,  .69,  .314}{17.8173} & \textcolor{red}{0.5338} & \textcolor{red}{0.5714} & \textcolor{red}{0.3874} & \textcolor[rgb]{ 0,  .69,  .314}{0.6263} & \textcolor{red}{0.3121} & \textcolor{red}{0.6667} & \textcolor{red}{0.5314} & \textcolor[rgb]{ 0,  .69,  .314}{0.8000}   & \textcolor{red}{0.1481} & \textcolor{red}{0.0909} \\
    \hline
    \end{tabular}%
  \label{tab:ensemble_result}%
\end{table*}%

Furthermore, we have also designed an experiment in which we combined the predictions of the top three teams for each task as if they were components of a single model. The ensemble strategy used for AMD classification and fovea localization tasks consisted on averaging the probabilities and coordinates submitted by each team, respectively; for disc and lesions segmentation tasks, we adopted a majority voting. Table~\ref{tab:ensemble_result} presents the quantitative results obtained for each task. Green values indicate that the ensembled result is better than the result achieved by the first team in the rank, while a red value indicates that it is worse. Notice that in the majority of cases, incorporating responses from the other models does not ensure an improvement, which emphasizes the importance of carefully designing the constitutive models of the ensemble.

\begin{table}[t]
  \centering
  \caption{AMD Classification (AMD and non-AMD classes) results stratified by disease stage.}
    \begin{tabular}{ccccc}
    \hline
    Teams & \makecell[c]{Early \\AMD\\ (24)} & \makecell[c]{Intermediate \\AMD \\ (9)} & \makecell[c]{Advanced \\AMD-\\dry (3)} & \makecell[c]{Advanced \\AMD-\\wet (53)} \\
    \hline
    VUNO EYE TEAM & \textbf{0.9159} & 0.9943 & 0.9861 & \textbf{0.9917} \\
    FobiddenFruit & 0.9090 & \textbf{0.9964} & 0.9743 & 0.9748 \\
    Zasti\_AI & 0.8901 & 0.9882 & \textbf{0.9914} & 0.9818 \\
    Muenai\_Tim & 0.8268 & 0.9877 & 0.97535 & 0.9809 \\
    ADAM-TEAM & 0.8284& 0.9775 & 0.9753 & 0.9632 \\
    WWW   & 0.8261 & 0.9578 & 0.9603 & 0.9502 \\
    XxlzT & 0.7661 & 0.9557 & 0.9700 & 0.9635 \\
    TeamTiger & 0.8312 & 0.9500 & 0.9539 & 0.9341 \\
    Airamatrix & 0.7186 & 0.9023 & 0.9855 & 0.9511 \\
    \hline
    \end{tabular}%
  \label{tab:AMD severity}%
\end{table}%

\subsection{Effects of disease stage and class imbalance}
\label{subsec:class-imbalance}

To evaluate the effect of different AMD stages on the binary AMD classification outputs of the teams, we mixed the samples of the corresponding AMD stage with non-AMD images and evaluated the resulting AUC values (Table \ref{tab:AMD severity}). As expected, the early AMD samples were more different to recognize than the intermediate, advanced dry, and advanced wet AMD cases, due to the lack of obvious pathological features. A potential solution could be to pay more attention to the difficult samples, which can be achieved by increasing their relative loss weight. 

In Table \ref{tab1} we can also see that the ADAM dataset is imbalanced in terms of the amount of samples with specific types of lesions. As the targets were subdivided into drusen, exudate, hemorrhage, scar and others, there are only a few positive samples per lesion type. The effect of this high class imbalance becomes apparent when we compare the lesion segmentation results with those obtained for optic disc segmentation and detection (Tables \ref{tab9} and \ref{tab7}, respectively). Despite the fact that the models are quite similar to one another (most of them based on U-shaped networks with high capacity encoders), the DICE values for optic disc segmentation obtained by all teams in the final stage were above 0.9 while those for lesion segmentation were mostly below 0.7. We hypothesize that this might be due to the few positive samples per lesion available in the dataset, apart from the difficulty of the problem itself (lesions are irregular in shape and sometimes not easy to distinguish one another). In subsequent studies, we will incorporate more images with lesions present to overcome this limitation. One potential way to alleviate this from a methodological perspective could be also to train the models at a patch level, which allows to artificially increase the relative amount of positive samples by oversampling patches from lesion areas.

\subsection{Incorporation of additional training datasets}
\label{subsec:additional-datasets}
\begin{table*}[htbp]
\setlength\tabcolsep{2pt}
  \centering
  \small
  \caption{Characteristics of the additional datasets used by the participating teams to train models for the four tasks.}
    \begin{tabular}{cccccc}
    \hline
    Additional datasets & Tasks & Teams & Annotators & Devices & Ocular diseases \\
    \hline
    ODIR  & AMD Classification & ForbiddenFruit, Airamatrix & -     & \makecell[c]{Canon, \\Zeiss,\\ Kowa} & \makecell[c]{diabetic retinopathy, glaucoma, \\cataract, AMD, pathologic myopia,\\ hypertensive retinopathy} \\
    REFUGE1 & \makecell[c]{Optic disc segmentation, \\ Fovea localization} & \makecell[c]{XxlzT, VUNO EYE TEAM, \\Zasti\_AI} & 7     & Zeiss & glaucoma \\
    IDRiD & \makecell[c]{Optic disc segmentation, \\ Fovea localization} & \makecell[c]{XxlzT, WWW, \\VUNO EYE TEAM, Voxelcloud} & -     & Kowa  & diabetic retinopathy \\
    RIGA  & Optic disc segmentation & WWW, VUNO EYE TEAM & 6     & Topcon, Canon & glaucoma \\
    PALM  & \makecell[c]{Optic disc segmentation, \\ Fovea localization} & VUNO EYE TEAM & 7     & -     & high myopia \\
    ARIA  & Fovea localization & Voxelcloud & -     & Zeiss & AMD, diabetic retinopathy \\
    DIARETDB1 & Lesion segmentation (exudate) & Airamatrix & 4     & Zeiss & diabetic retinopathy \\
    fundus 10k & Lesion segmentation (scar) & Airamatrix & -     & -     & diabetic retinopathy\\ 
    \hline
    \end{tabular}%
  \label{tab:additional data intro}%
\end{table*}%
One key factor in supervised deep learning is the requirement of abundant and well-labelled training data, as it allows training larger, more accurate and robust models. If not enough data is available, the capabilities of a neural network to generalize well to other unseen images is frequently diminished, reducing its potential to become a clinically relevant tool~\citep{shorten2019survey, geirhos2020shortcut}. We observed that several teams decided to incorporate additional training data from other publicly available datasets based on the assumption that this extra samples might improve the performance of their trained models.

Table~\ref{tab:additional data intro} shows the additional datasets used by the participating teams in our challenge. The table lists devices, ocular diseases involved, and number of annotators, the three key factors that may influence the performance of machine learning models. Notice that the types of ocular diseases included in these datasets sometimes go beyond AMD, including diabetic retinopathy, glaucoma, cataracts, pathologic myopia, hypertensive retinopathy, etc. With respect to the labels, we observed that half of the additional datasets did not provide information about the number of annotators involved in producing the ground truth labels. If the targets were annotated by single observer, it is more likely that the references contain a certain amount of error, which can further affect the models. Using labels produced by ensembling responses from multiple readers, as performed when creating the ADAM dataset, might reduce human bias and ensure more trustworthy labels. Furthermore, relying on the clinical records at the moment of creating these annotations makes it more likely to achieve a high quality of the reference standard.

\begin{figure}[!t]
\centering
\includegraphics[width=0.8\linewidth]{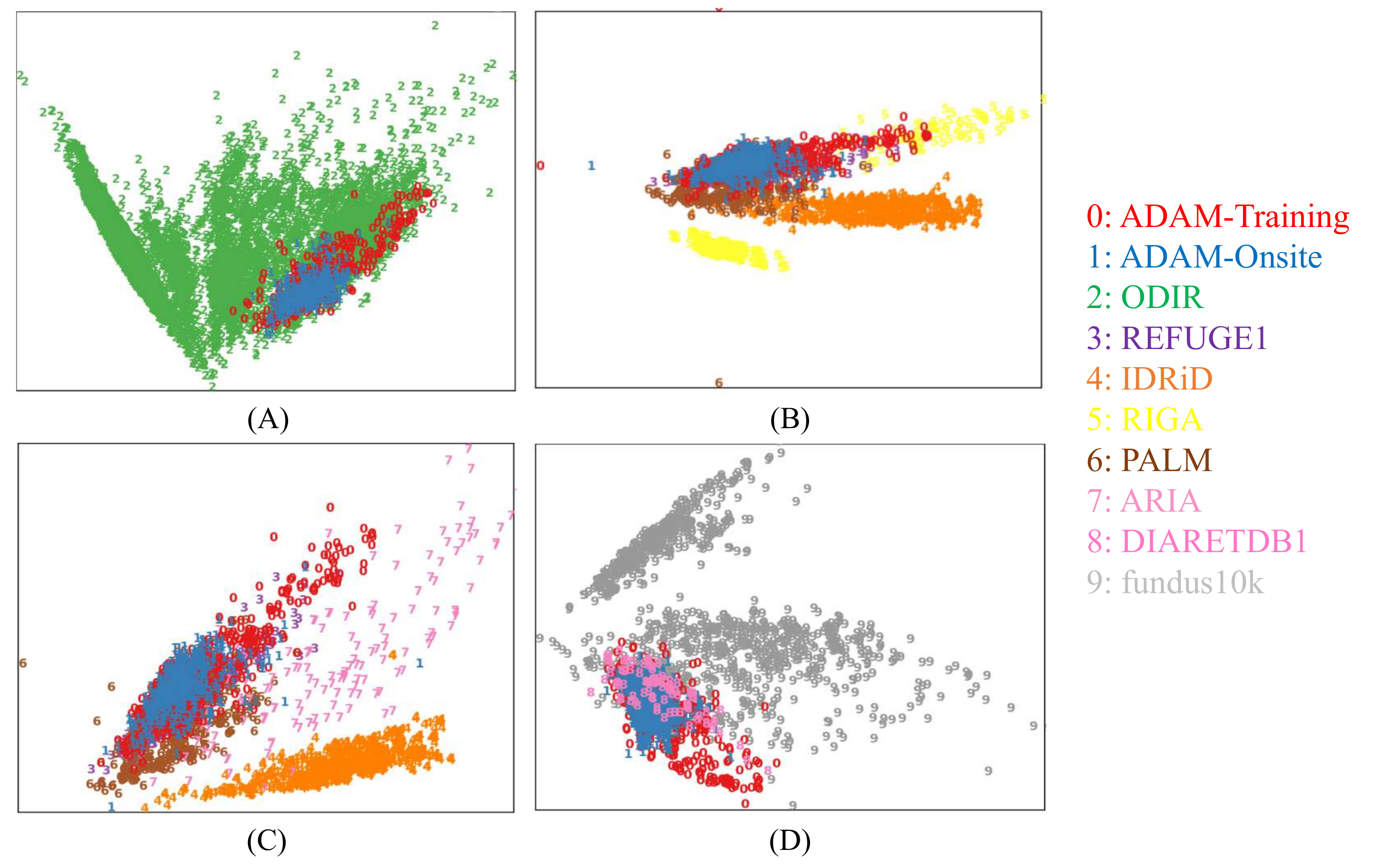}
\caption{t-SNE~\citep{van2008visualizing} representation of the original images in the datasets used by the teams to increase their training data, separated per task. (A) Classification of AMD and non-AMD; (B) Optic disc detection and segmentation; (C) Fovea localization; (D) Lesion detection and segmentation.}
\label{fig:datadistribution}
\end{figure}

\begin{figure}[t]
\centering
\includegraphics[width=0.75\linewidth]{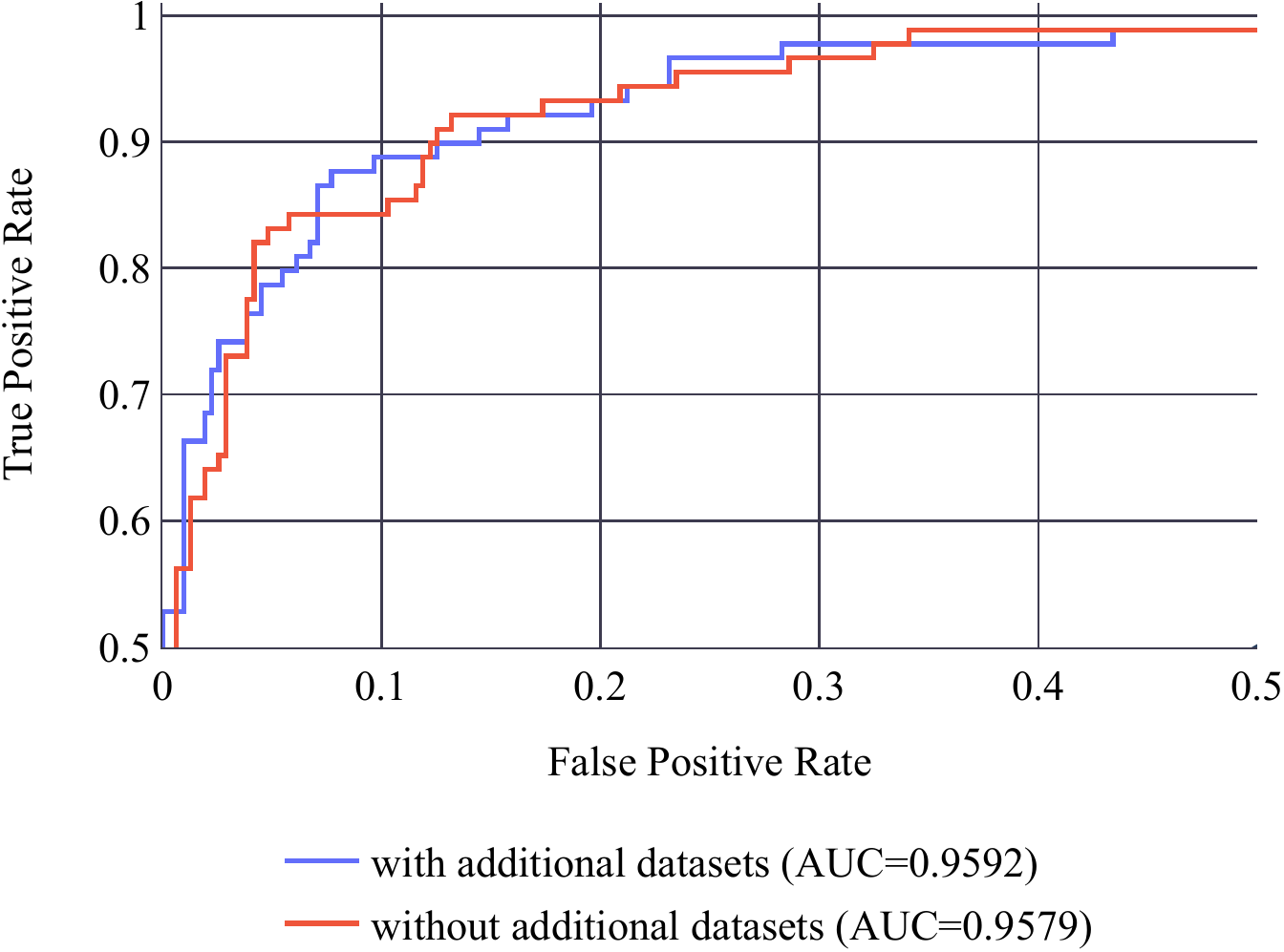}
\caption{ROC curves and AUC values obtained by the Forbiddenfruit team model for AMD classification in the onsite dataset, training with and without additional images from ODIR.}
\label{fig24}
\end{figure}

\begin{figure}[t]
\centering
\includegraphics[width=0.8\linewidth]{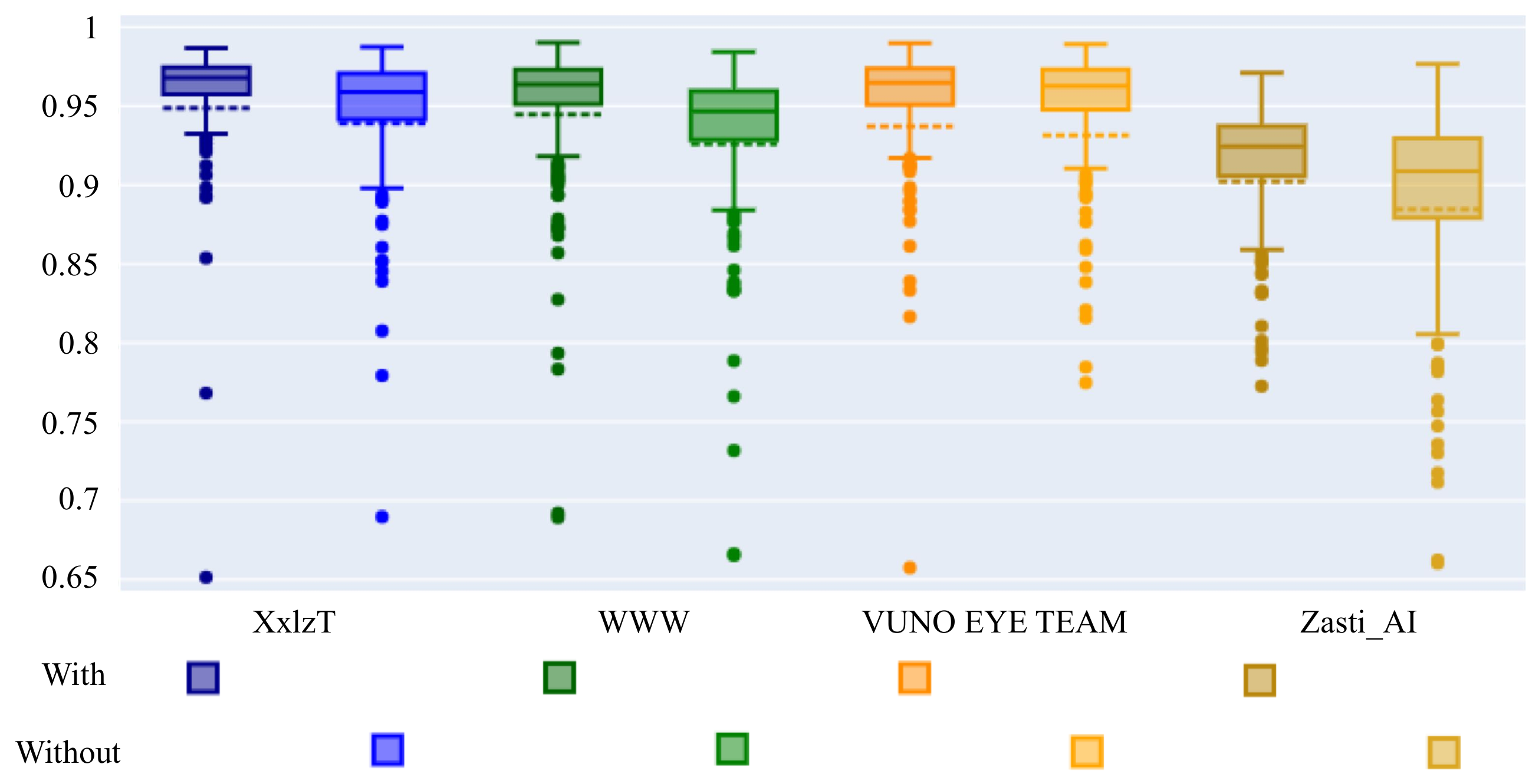}
\caption{Distribution of Dice values obtained by XxlzT, WWW, VUNO EYE TEAM and Zasti\_AI for optic disc segmentation in our onsite set, with (dark colors) and without (light colors) incorporating additional datasets for training.}
\label{fig25}
\end{figure}

\begin{figure}[!t]
\centering
\includegraphics[width=0.8\linewidth]{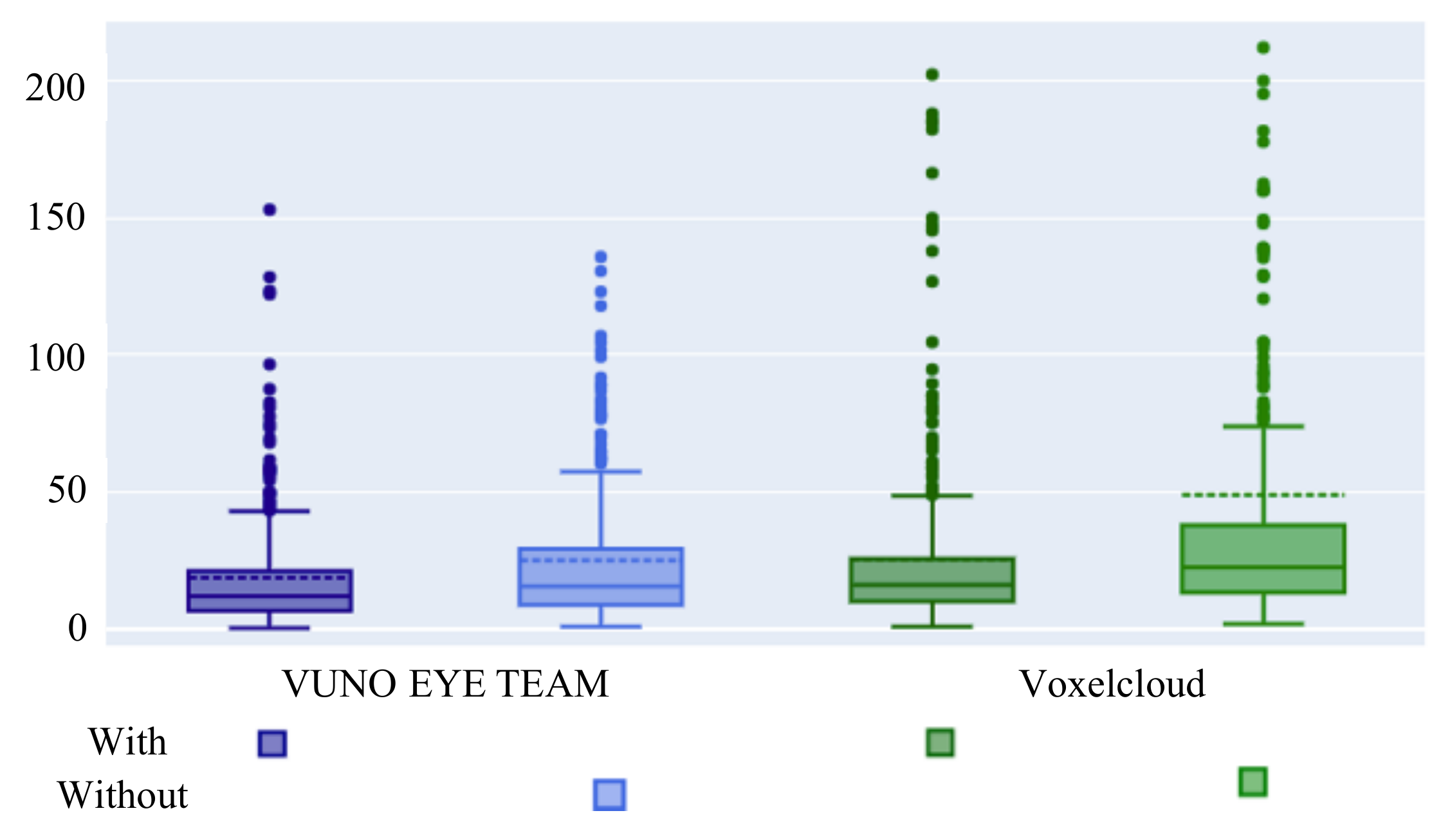}
\caption{The ED evaluation results of fovea localization predicted by the models trained with and without additional datasets by VUNO EYE TEAM and Voxelcloud team.}
\label{fig26}
\end{figure}

Fig.~\ref{fig:datadistribution} depicts the t-SNE distribution of the original images in the datasets listed in Table~\ref{tab:additional data intro}, separated by each of the challenge tasks. The t-SNE was implemented via scikit-learn package~\citep{pedregosa2011scikit}, which is an open source machine learning toolkit base on Python. Specifically, the original images were resized to $64\times64$ pixels, and were flattened to 4096-dimensional feature, then the operation of dimension reduction from 4096 to 2 was implemented by t-SNE in scikit-learn package with the dimension of the embedded space setting to 2, the initialization of embedding pattern setting to ‘pca’, and the other parameters setting to default value. From the figure, it can be seen that the additional images considered by some teams indeed expand the distribution range of the training data, which might improve generalization. However, the distribution of the onsite set is similar to that of the original training set, so it is difficult to actually quantify the generalization ability of the models on it. In the AMD classification task, for instance, the ForbiddenFruit team used ODIR as a complementary dataset\footnote{\url{https://odir2019.grand-challenge.org/dataset/}}. We evaluated their results with and without using these additional images, and depicted the resulting ROC curves in Fig.~\ref{fig24}. Notice that a slight improvement in the AUC values is observed. For optic disc segmentation, teams XxlzT, WWW, VUNO EYE TEAM and Zasti\_AI also used additional datasets, namely IDRiD~\citep{porwalIDRiDDiabeticRetinopathy2020a}, REFUGE~\citep{orlandoREFUGEChallengeUnified2020a}, RIGA~\citep{almazroaRetinalFundusImages2018}, and PALM~\citep{PALM}. Fig.~\ref{fig25} shows the distribution of Dice values obtained by the four teams using models trained with and without the additional datasets. If the extra data is not included for training their models, results achieved by XxlzT, WWW, VUNO EYE TEAM, and Zasti\_AI decrease by 1.02\%, 1.98\%, 0.61\%, and 1.94\%, respectively.  The same holds for fovea localization, where VUNO EYE TEAM and VoxelCloud used public datasets (including IDRiD, REFUGE, PALM, ARIA~\citep{chea2021classification}), with Voxelcloud furthermore incorporating a proprietary dataset. Fig.~\ref{fig26} shows that the ED error metric obtained by VUNO EYE TEAM and Voxelcloud team without using additional data are increased by 34.37\% and 93.54\%, respectively, compared to those obtained using extra images. Finally, Table \ref{tab10} shows Dice and $F_1$ scores obtained by Airamatrix for segmenting and detecting exudates and scares, with and without using images from DIARETDB1\footnote{\url{https://www.it.lut.fi/project/imageret/diaretdb1/}} and fundus10k\footnote{\url{https://github.com/li-xirong/fundus10k}}. While Dice values are only slightly increased, the effect of this additional data over the $F_1$ detection metric is much more evident, especially for scar segmentation. 
In summary, we can see that incorporating more images have a generally positive effect on the performance, however, this cannot be directly extrapolated to an improvement in generalization. Future challenges aiming to effectively measure this behavior would need to incorporate multiple test sets from multiple institutions and devices and with different comorbidities~\citep{fang2022refuge2, wu2022gamma}.

The experimental results indicate that on the challenge onsite set, except for the fovea location task, the use of additional datasets did not sharply improve the performances of the models. The use of additional datasets can expand the distribution of the training data, as shown in Fig.~\ref{fig:datadistribution}, however, the data distribution of our onsite set was similar to that of the training set, so the results on this set do not reflect the out of sample generalization ability of the models. In addition to enriching the distribution of the existing training set, it can be seen from Fig.~\ref{fig:datadistribution} that the additional datasets may complement the amount of the samples in the distribution of the existing training set. These samples may be helpful to improve the predictions on the samples with such data distribution. However, for AMD classification task and optic disc segmentation task, the existing original training data is enough for the model to find the optimal parameters, so the additional data plays a small role. In contrast, these additional data are important for the fovea localization task, because it is a regression task, which is more sensitive to the loss value during the training process than the classification task. For example, the prediction probabilities of 0.7 and 0.99 in the classification task can be regarded as the same category, but there would be a 0.29 error in the regression task. Hence, compared with the classification task, the regression task benefits from larger sample sizes. This explains our experimental conclusion that in the fovea localization task, the model trained with additional data sets achieved a noticeable improvement. For the task of lesion detection and segmentation, there were very few positive samples in the original training set. Theoretically, adding additional annotated data would lead to a substaintal improvement in the model performance, but the experimental results (Table~\ref{tab10} ) did not support this. We speculated that the main reason was that different datasets had inconsistent labeling standards for lesions. Therefore, in the future work, we need to collect a large size of widely distributed data, collected by different devices, with uniform and consistent labeling standards, and make better clinical characterization of them, to provide support to study algorithms with strong generalization ability and to study diagnostic methods for cases with multiple ocular disorders. 


\subsection{Clinical domain knowledge incorporation}
\label{subsec:clinical-knowledge}

\begin{table}[t]
\small
    \centering
    \caption{Dice and $F_1$ values obtained by Airametrix team with and without additional datasets for segmentation and detection of exudates and scars in our onsite dataset.}
    \begin{tabular}{c|c|c|c|c}
    \hline
    \multirow{2}{*}{Additional datasets?} & 
    \multicolumn{2}{c|}{Exudate} & 
    \multicolumn{2}{c}{Scar}\\  
    \cline{2-5}
    ~ &Dice&  $F_{1}$ &  Dice &$F_{1}$  \\
    \hline
    With&     \textbf{0.2606}&  \textbf{0.4673}&   \textbf{0.4080} & \textbf{0.6500} 
    \\               
    \hline
    Without&  0.2574&  0.4545&  0.4006&  0.6047  
    \\        
    \hline
    \end{tabular}
\label{tab10} 
\end{table}
Although the approaches used by the participating teams were all based on deep learning frameworks, many teams incorporated  prior clinical domain knowledge for some specific tasks. For example, the Airametrix team designed a joint learning method for optic disc segmentation and fovea localization that simultaneously takes advantage of the position of the disc and the macula area. Such an approach ranked first for optic disc segmentation, which in principle might indicate that this extra supervision plays a relevant role for the task. For fovea localization, VUNO EYE TEAM (ranked first) and the Voxelcloud team (ranked third) used prior knowledge about the shape of the vessels, based on the assumption that this may help to better identify the right fovea location. The ForbiddenFruit team, on the other hand, incorporated the AMD/non-AMD probability into its lesion segmentation model, as both tasks are closely clinically related. The high rankings achieved by the teams that made use of prior clinical domain knowledge might be taken as a good indicator of the positive effect it brings toward learning more accurate models. From a methodological perspective, we believe that future AMD detection models should incorporate information about the size and location of lesions with respect to the grid and standard circles used by ophthalmologists, which may not only improve their performance but also produce more interpretable and trustworthy solutions.

\subsection{Clinical implications of the results}
\label{subsec:clinical-implications}

Automated optic disc detection and segmentation, fovea localization and AMD-related lesion detection and segmentation allow clinicians to obtain a holistic description of a fundus image that allows better and more efficient diagnostics of AMD. Having accurate models for each of those is therefore crucial to ensure a proper characterization of AMD and improve clinical outcomes. The ADAM challenge was designed with the purpose of evaluating in a single, uniform framework a series of methodological proposals for tackling these tasks. For optic disc detection and segmentation, the best performance was obtained by the XxlzT team, with a Dice of 0.9486 and a $F_{1}$ score of 0.9913. While the $F_1$ value indicates an almost perfect ability to identify if the optic disc is visible on a given fundus image, the segmentation results are similar to those quantified before in the first edition of the REFUGE challenge~\citep{orlandoREFUGEChallengeUnified2020a}, where the top ranked team achieved an average Dice value of 0.9602.
For fovea localization, the winning team VUNO EYE TEAM obtained an average ED of 18.6 pixels, which is about one-tenth of the radius of the optic disc. These accurate results provide good insights about the current status of deep learning models for providing quantitative information about fundus structures. 

\begin{table}[htbp]
  \centering
  \caption{The comparison of the manual labels annotated by 7 ophthalmologists and the reference standard in the tasks of optic disc segmentation and fovea localization.}
    \begin{tabular}{cccc}
    \hline
          & \multicolumn{2}{c}{Disc Segmentation} & Fovea Localization \\
    \cline{2-4}
          & $F_{1}$    & DICE  & ED \\
    \hline
    user1 & 0.9982 & 0.8685 & 30.8106 \\
    user2 & 1.0000 & 0.8991 & 28.0984 \\
    user3 & \textcolor[rgb]{ 1,  0,  0}{0.9965} & \textcolor[rgb]{ 1,  0,  0}{0.7469} & \textcolor[rgb]{ 0,  .69,  .314}{21.1393} \\
    user4 & 1.0000 & 0.8979 & 26.3184 \\
    user5 & 0.9982 & 0.9068 & \textcolor[rgb]{ 1,  0,  0}{62.8048} \\
    user6 & 0.9982 & 0.8553 & 29.9480 \\
    user7 & \textcolor[rgb]{ 0,  .69,  .314}{1.0000} & \textcolor[rgb]{ 0,  .69,  .314}{0.9439} & 46.5350 \\
    \hline
    \end{tabular}%
  \label{tab:users_labelling}%
\end{table}%

Table \ref{tab:users_labelling} shows a quantitative evaluation of the manual annotations made by each of the 7 independent ophthalmologists with respect to the final reference standard, for the 400 images on the onsite set. For optic disc segmentation, all the optic discs involved in 400 images were correctly detected, and the highest average DICE value reached was 0.9439. We can see that the automated method proposed by XxlzT and Airamatrix teams obtained better results in average than the best reader, and that all the finalist teams produced segmentation with an average DICE higher than the one obtained by the worst performing observer. A similar behavior can be seen for the fovea localization task, where the automated methods proposed by VUNO EYE TEAM and ForbiddenFruit teams (see Table~\ref{tab8}) achieved better results than the best annotator, and the automatic methods proposed by the top six teams of the finalists produced better localization results than the worst annotator. In summary, this also supports our observation that machine learning models are mature enough to achieve equivalent or even more accurate results than those manually obtained by ophthalmologists for the optic disc segmentation and fovea localization.

The lesion detection and segmentation results, on the other hand, are far from achieving the accuracy required for clinical applications yet. However, it is not possible to ensure that this is due to methodological failures, as the small training sample size may have played a harmful effect on the results. Future challenges should incorporate a larger amount of images with lesions. Nevertheless, it must be pointed out that having accurate, pixel-wise manual annotations of pathological areas is extremely costly, especially if they are to be produced by multiple annotators. 

Finally, it is worth mentioning that best performing team on the classification task (VUNO EYE TEAM) achieved an AUC of 0.9714, which may be taken as a positive sign towards having tools for automated screening for the disease. Further improvements are still needed before their clinical application, though, especially concerning their interpretability, calibration and generalization ability to multiple ethnicities and camera vendors. These three points were not explicitly evaluated in the ADAM challenge, and should be taken into account in the future challenges.

\section{Conclusion}   
In this paper, we summarized the methods, results and main findings of the first international contest on AMD detection from fundus images, the ADAM challenge. We analyzed and compared the performances of the 11 teams that participated in the onsite edition of the ADAM challenge at ISBI 2020, on four AMD-related tasks. The remarkable results achieved for optic disc detection and segmentation, fovea localization, and classification of AMD disease are good indicators of the maturity level of the automated solutions for these tasks. Furthermore, we observed that the ensemble strategy utilized by many teams is in line with observations made in other challenges~\citep{orlandoREFUGEChallengeUnified2020a}, and is an easy to implement solution to improve performance. Incorporating clinical prior knowledge such as the location of blood vessels, the optic disc, and the fovea or the disease classification label, on the other hand, was observed to aid to achieve better results for certain tasks. A similar behavior was observed when adding extra datasets for training, although the improvements were not as notorious for tasks such as AMD classification. However, since the distribution of the onsite set is similar to that of the original training set, it is not possible to use this set as an indicator of improved generalization ability.

We encourage future research in this field to take a more holistic approach, and to simultaneously analyze fundus structures and AMD-related biomarkers. To contribute to this direction, our large dataset with comprehensive labels of fundus structures, AMD lesions and disease categories, as well as our evaluation framework, are made publicly accessible through the website at \url{https://amd.grand-challenge.org/}. As medical data is sensitive and subject to particularly strict accessibility rules, researchers interested in using the dataset are required to sign up to the ADAM challenge by registering their personal information (name, institution and e-mail) for participation approval. After the approval, the researchers can access the data freely at  \url{https://amd.grand-challenge.org/download/}. Participants are invited to utilize the dataset to develop more robust and novel algorithms for AMD screening from fundus images, and to benchmark their performance by submitting their results to the ADAM website.\\

\section{Appendix. Summary of Challenge Solutions} 
\label{appendix}
This appendix provides detailed descriptions of the methods of the participating teams in four tasks.

\subsection{Classification of AMD and non-AMD images task}
\label{appendix_A}

\begin{figure}[t]
\centering
\includegraphics[width=1\linewidth]{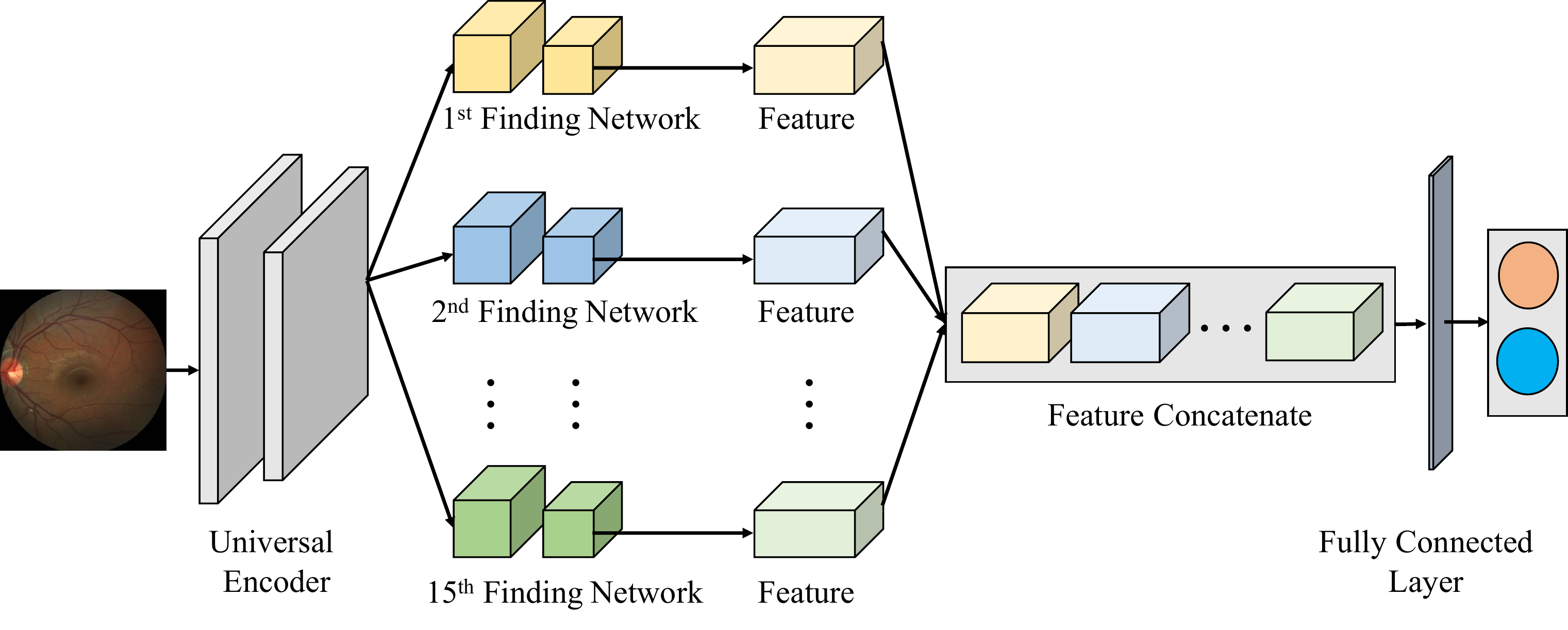}
\caption{The framework of the VUNO EYE TEAM for AMD classification, where 15 finding networks were utilized to extract the image features.}
\label{fig4}
\end{figure}








A brief summary of the methods adopted by the participating teams is shown in Table~\ref{tab2} in the main manuscript.
The top 5 teams in this classification task utilized the ensemble method. Among them, the 1st and the 4th teams adopted a self-ensemble strategy, which meant that the integrated models had the same structures but different parameters. In particular, the VUNO EYE TEAM (1st) trained an EfficientNet~\citep{tanEfficientNetRethinkingModel2019} by using a dataset with 15 types of lesion labels (hemorrhage, hard exudate, cotton wool patch, drusen, retinal pigmentary change, vascular abnormality, membrane, fluid accumulation, chorioretinal atrophy, choroidal lesion, myelinated nerve fiber, retinal nerve fiber layer defect, glaucomatous disc change, non-glaucomatous disc change, and macular hole), and obtained 15 models~\citep{sonDevelopmentValidationDeep2020}. Then, they obtained the AMD classification results by combining the feature maps from the 15 models and using a fully connected layer and Sigmoid activation function (as shown in Fig.~\ref{fig4}). The Muenai\_Tim team (4th) also used the EfficientNet architecture that has been widely successful at computer vision fields and optimal for memory consumption~\citep{tanEfficientNetRethinkingModel2019}. They realized the self-ensemble by integrating the models which have the local minimal loss during the training processing.

\begin{figure}[t]
\centering
\includegraphics[width=1\linewidth]{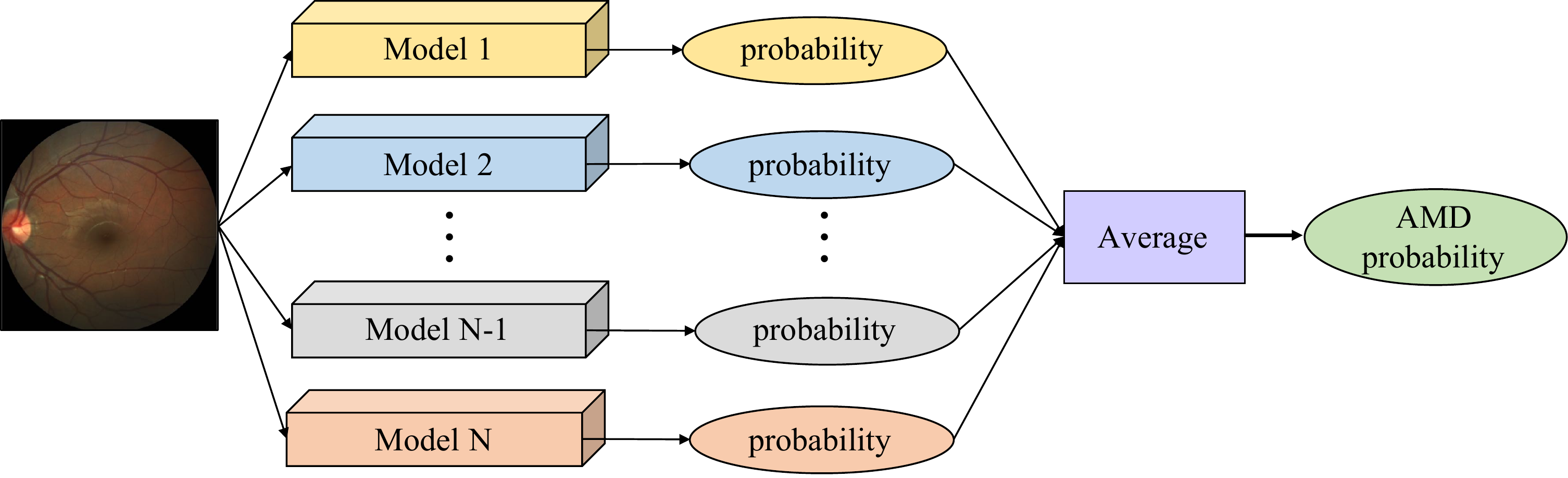}
 \caption{The average ensemble framework of the Zasti\_AI team and the ADAM-TEAM for AMD classification task.} 
\label{fig5}
\end{figure}

The other three teams took advantage of the ensemble approach of multiple models. The main difference in these approaches is the fusion strategy. A common ensemble strategy is averaging. The Zasti\_AI team (3rd) and the ADAM-TEAM (5th) both adopted the average strategy to ensemble the performances of the multiple standard deep learning models (as shown in Fig.~\ref{fig5}). The Zasti\_AI team built 4 models with the EfficientNet architectures~\citep{tanEfficientNetRethinkingModel2019}, 2 models with the ResNext structures~\citep{xieAggregatedResidualTransformations2017}, 1 model with Inception-ResNet~\citep{szegedyInceptionv4InceptionResNetImpact2017}, and 1 model with SENet architecture~\citep{huSqueezeandExcitationNetworks2018}. The ADAM-TEAM built 4 models with Inception-v3, Xception, ResNet-50~\citep{heDeepResidualLearning2016} and DenseNet-101 architectures, respectively. In these two methods, the results of each model contributed the same to the final prediction results. However, the ForbiddenFruit team (2nd) designed a novel ensemble function, which made the results of different models contribute differently. They ensembled 5 models, which included EfficientNet-B2, -B4, -B3, -B7~\citep{tanEfficientNetRethinkingModel2019} and DenseNet-201~\citep{huangDenselyConnectedConvolutional2017}, by using the following function: 
\begin{equation}
\varepsilon(I)=\exp(\frac{\sum_{i=1}^{N}\omega_{i}log\mathcal{A}_{i} (I)}
{\sum_{i=1}^{N}\omega_{i}})
\end{equation}
where $I$ is the input image, $\mathcal{A}_{i}$ represents the i-th models, and $\omega_{i}$ (positive integer) are ensemble weights. The number $N$ is 5, and the weights $\omega_{i}$ were found experimentally: 3 for EfficientNet-B2, 2 for EfficientNet-B4, 1 for the other three models.  

\begin{figure}[t]
\centering
\includegraphics[width=0.7\linewidth]{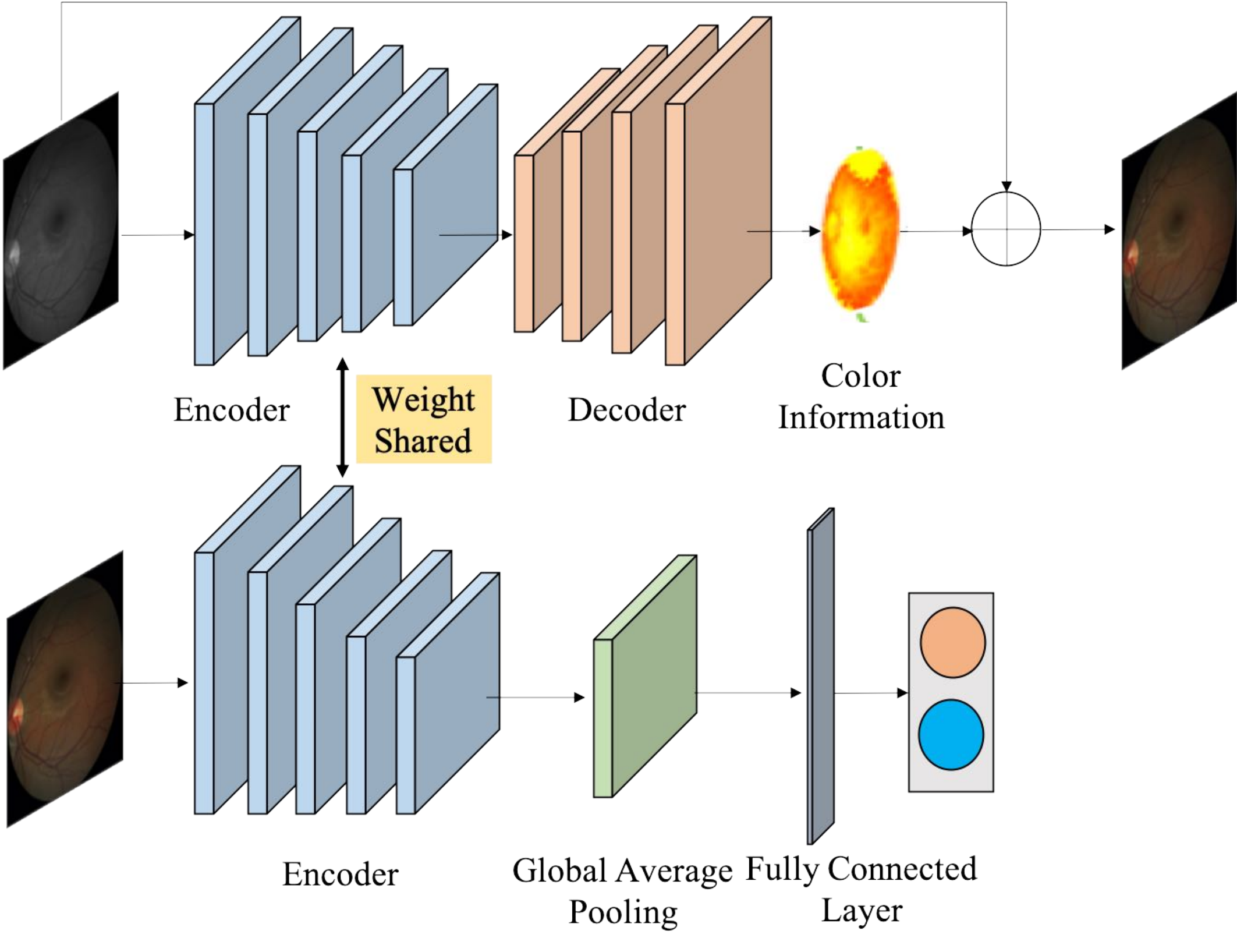}
\caption{The framework of the XxlzT team for AMD classification, where the two encoder share the same parameters.}
\label{fig6}
\end{figure}

The remaining four teams designed their frameworks without ensemble strategy. The TeamTiger team (8th) directly utilized ResNet101 to realize the classification. To obtain the discriminative feature maps for better classification, the other teams developed several strategies to improve the feature extraction modules. The WWW(6th) and the Airamatrix (9th) teams pre-trained an EfficienNet-B7 and an EfficientNet-B4 based on ImageNet respectively, and then fine-tuned them based on the clinical dataset. The XxlzT team (7th) proposed a self-supervised module to obtain the parameters of the encoder architecture, which is used to extract the image features. As shown in Fig.~\ref{fig6}, they extracted the grayscale image from the fundus image and then used an encoder-decoder framework to generate the color information, which was then superimposed with the grayscale image to generate the fundus image. In the classification task, they utilized the encoder architecture and the shared parameters to extract features from the fundus images and then used a fully connected layer to realize the classification.  

In addition, the training datasets of the 7 teams were completely derived from the ADAM dataset, while that of the Forbiddenfruit and Airamatrix teams contained ODIR dataset\footnote{\url{https://odir2019.grand-challenge.org/dataset/}}. The loss functions adopted by all teams are based on cross-entropy (CE), in which the WWW team used weighted cross-entropy loss~\citep{cuiClassBalancedLossBased2019}  to solve the data imbalance problem.  

\subsection{Detection and segmentation of optic disc task}
\label{appendix_B}

The methods used by the participating teams are summarized in Table~\ref{tab3} in the main manuscript. Since this task involved detection and segmentation of the optic discs, three teams adopted a multi-step training strategy, which first classifies the fundus images into with or without complete optic disc, and then segments the complete optic disc region. For classification, the XxlzT team (2nd) utilized ResNet, the WWW team (4th) and the Muenai\_Tim(8th) team used EfficientNet-B7 and -B0. In the segmentation process, all three teams adopted U-Net~\citep{ronnebergerUNetConvolutionalNetworks2015}, while for more precise results, the XxlzT team further performed fine segmentation of optic disc using a DeeplabV3 network~\citep{chenRethinkingAtrousConvolution2017} on the previous segmentation results, and the Muenai\_Tim team used an emsemble of the U-Net variant, U-Net++, and EfficientNet-B7. 

\begin{figure}[t]
\centering
\includegraphics[width=0.8\linewidth]{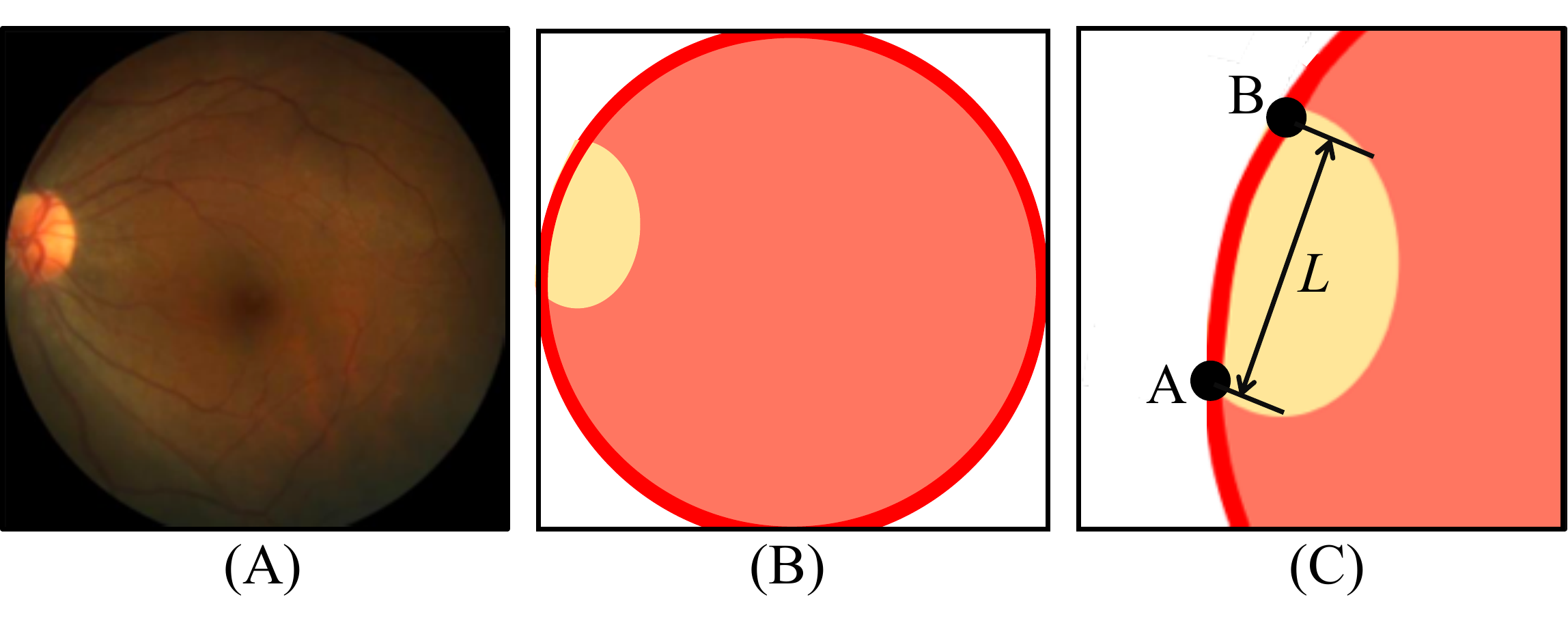}
\caption{The rule for estimating integrity of optic disc designed by the ForbiddenFruit team. (A) fundus image, (B) the camera’s field of view (red) and optic disc segmentation (yellow), (C) enlarged figure of the optic disc region and the intersecting length $L$ of the segmented optic disc and the edge of the camera’s field of view.}
\label{fig8}
\end{figure}

Half of the participating teams utilized the strategy of segmenting the optic disc directly. Among them, three teams adopted U-Net. The differences between these three methods are the encoder structure of the U-Net and the ensemble strategy. Specifically, the TeamTiger team (6th) and the Zasti\_AI team (8th) respectively adopted EfficientNet-B7 and ResNet architectures as the encoders. Moreover, the Zasti\_AI team used an adversarial training setting~\citep{shankaranarayanaJointOpticDisc2017} to improve the accuracy of segmentation results. The ADAM-TEAM (7th) respectively used Inception-v3, ResNet-50, EfficientNet-B3, and DenseNet-101 as encoders in the U-Net based architectures. And, they determined the segmentation results by averaging the predictions of these four different models. Except for the U-Net architecture, the ForbiddenFruit team (3rd) utilized two Feature Pyramid Networks (FPN)~\citep{linFeaturePyramidNetworks2017}, one of which used EfficientNet-B0 as encoder, and another used EfficientNet-B2. The outputs of the two FPNs were then averaged and the probability map was thresholded at 0.5. The CHING WEI WANG(NTUST) (10th) team adopted the segmentation modules, which is based on a fully convolutional networks (FCN) with VGG16 as encoder, in the AI Explore platform. After segmentation, these five teams designed several post-processing steps to remove the incomplete optic disc. For example, the ForbiddenFruit team determined whether the optic disc was intact by calculating the degree of intersection between the segmentation results and the camera's field of view. It can be seen in Fig.~\ref{fig8}, if the length of intersection $L > 5\%$ of the height in the segmentation mask, the mask should be removed. The Zasti\_AI team simply kept a threshold on the area largest connected component in the segmentation map during the post-processing stage and discarded if the area was smaller than the threshold.

\begin{figure}[t]
\centering
\includegraphics[width=1\linewidth]{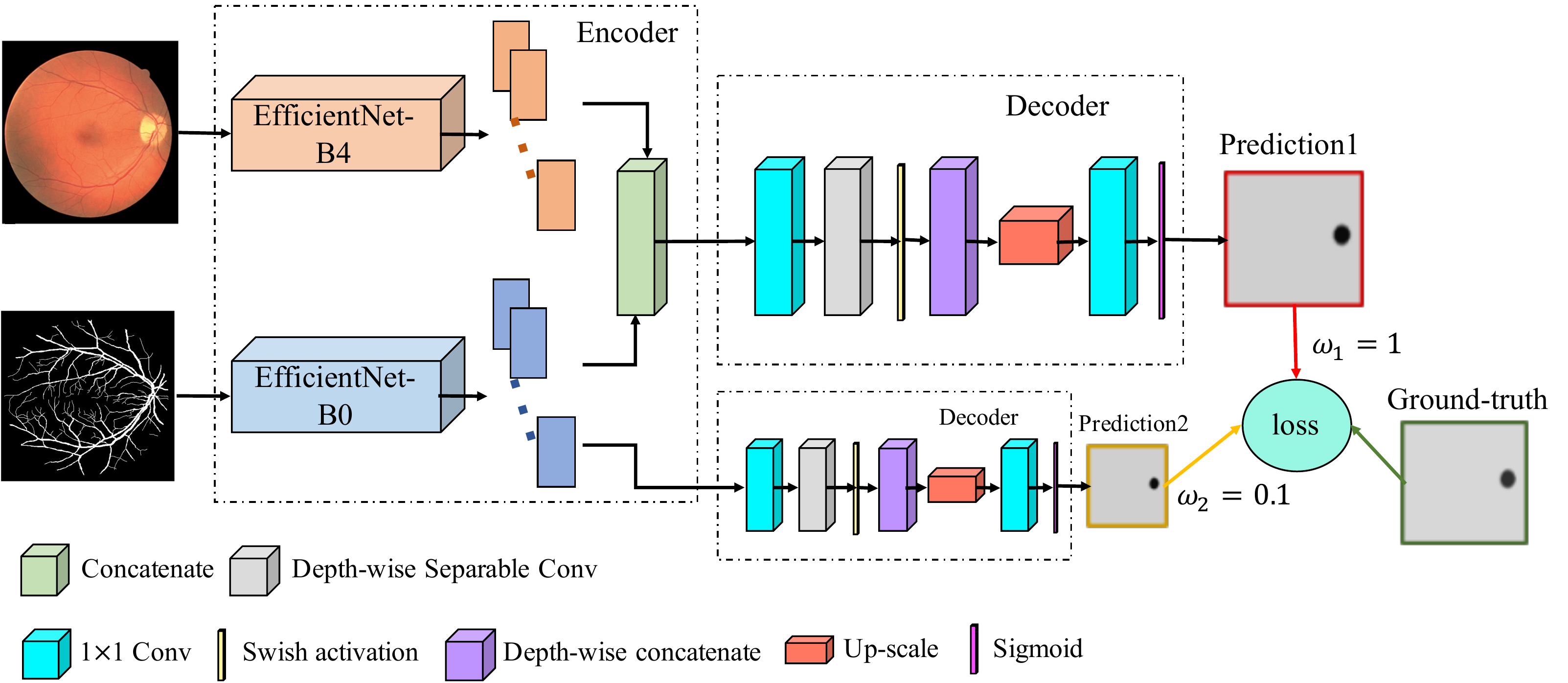}
\caption{The framework of the VUNO EYE TEAM for the optic disc segmentation task, which using vessel segmentation mask as input simultaneously.}
\label{fig.vuno}
\end{figure}

Notably, two teams considered other clinical information and used the multi-task strategy to segment the optic disc. Specifically, the Airamatrix team (1st) designed a framework to solve both the optic disc segmentation and fovea localization tasks. They transferred the localization task to a segmentation task, so joint training strategy was adopted to make the model simultaneously segment the optic disc and fovea regions. 
The FCNs with the ResNet-50 as backbone were used for optic disc and fovea segmentation. In addition, they applied erosion on the detected masks to further improve the accuracy. The VUNO EYE TEAM (5th) incorporated the blood vessels during training. They took a fundus image and a vessel image as input. The network consisted of two branches (see Fig.~\ref{fig.vuno}). 
One branch, which was EfficientNet-B4, processed the fundus image, and the other branch, which was EfficientNet-B0, operated on the vessel image. The penultimate feature maps of the fundus branch were concatenated to those of the vessel branch. In the decoder module, they up-scaled the feature maps using 1$\times$1 convolutions, depth-wise separable convolutions~\citep{howardMobileNetsEfficientConvolutional2017}, swish activation functions~\citep{ramachandranSearchingActivationFunctions2017}, and depth-wise concatenation and then scaled the feature maps to yield those with the same size of the input. The final segmentation layer is generated using a 1$\times$1 convolution followed by a Sigmoid function. They imposed a loss weight of 0.1 on the vessel branch as the vessel shape can give a good indication of the optic disc region, and the loss weight of the last layer was set to 1. In addition, the VUNO EYE TEAM used the snapshot ensemble approach to integrate the models obtained from epoch 62, 77, 93, 109, and 124 during training.

In this task, half of the participating teams used additional datasets. In detail, three teams (XxlzT, VUNO EYE TEAM, and Zasti\_AI) used REFUGE dataset~\citep{orlandoREFUGEChallengeUnified2020a}. Three teams (XxlzT, WWW, and VUNO EYE TEAM) used IDRiD dataset~\citep{porwalIDRiDDiabeticRetinopathy2020a}. Two teams (WWW and VUNO EYE TEAM) used RIGA dataset~\citep{almazroaRetinalFundusImages2018}. The VUNO EYE TEAM also used PALM dataset~\citep{fuIChallengePALMPAthoLogicMyopia2019}. In the selection of the loss function, since the segmentation task can also be regarded as a binary task, eight teams used 
cross-entropy loss. In addition, the ADAM-TEAM and the Muenai\_Tim team used Dice Loss, the ForbiddenFruit team used Dice Loss and Focal Loss, and the TeamTiger team used Jaccard Loss.

\subsection{Localization of fovea task}
\label{appendix_C}








Table~\ref{tab4} of the main manuscript shows the summary of the methods of the participating teams in this task. 
The methods based on the regression network were to predict the x and y coordinates directly. The TeamTiger team (8th) and the Muenai\_Tim team (7th) directly used EfficientNet-B7 to predict the coordinates. While before regression, the Muenai\_Tim team used EfficientNet-B0 to determine whether the fundus image contained fovea, and the coordinate was set to (0, 0) if it did not. 

The Zasti\_AI team (5th) created distance maps using the Euclidean distance transforms, and converted the coordinate regression problem into an image generation problem, which was to generate the distance map from the fundus images using a generative adversarial network~\citep{shankaranarayanaJointOpticDisc2017}. Finally, they clustered one percent of the highest intensities and segment out the largest cluster. The fovea coordinate was the centroid of this largest cluster. 

The VUNO EYE TEAM (1st) designed a segmentation mask of a single-pixel of the fovea, and two deviation masks on the x and y axis to deal with the inconsistent size of the output features and the input images. Then, they used U-Net to predict the above three masks. Specifically, they utilized the same network architecture designed in task 2, except for the last layer which consisted of a confidence map, a map for x-offset, and a map for y-offset. Similar to the loss function in task 2, they also gave losses to the offset maps, which are generated from the final feature maps of the vessel branch. 

In addition to the coordinate information, seven teams considered the region information of the fovea. Five teams of them trained segmentation networks guided by the binary fovea masks, where the fovea being foreground was represented by a circle or a box centered at fovea with a radius according to the optic disc radius, the image height, or fixed size. Two teams transferred the coordinate prediction problem to an object detection problem with marking the detection box label according to the given coordinates. In detail, the CHING WEI WANG(NTUST) team (8th) transformed the fovea location $(x, y)$ as a $100 \times 100$ pixels box, and utilized FCN module in AI Explore platform to segment the box region. The center of the output box was the location of the fovea. The FrobiddenFruit team (2nd) designed a regression branch and a segmentation branch. In the regression branch, three models were built respectively by VGG-19, Inception-v3, and ResNet-50. 
Besides, an additional Inception-v3 model was trained using a more realistic default location when the fovea is invisible: (1.25, 0.5). When the latter model disagreed with the others by more than 0.5 along x coordinate, the fovea was considered invisible. Otherwise, the predictions of the above four models were averaged to be the final regression result. In the segmentation branch, a circle centered on the fovea, with a diameter equal to 5\% of the image height, was used as ground truth, and three FPNs with EfficientNet-B0, -B1, and -B2 as encoders were trained. 
Finally, images were first processed by the segmentation branch. If a fovea was detected, then the centroid was used as fovea location prediction. If no fovea was detected, then the image was processed by the regression branch for less precise but more robust estimation. 

\begin{figure}[t]
\centering
\includegraphics[width=0.8\linewidth]{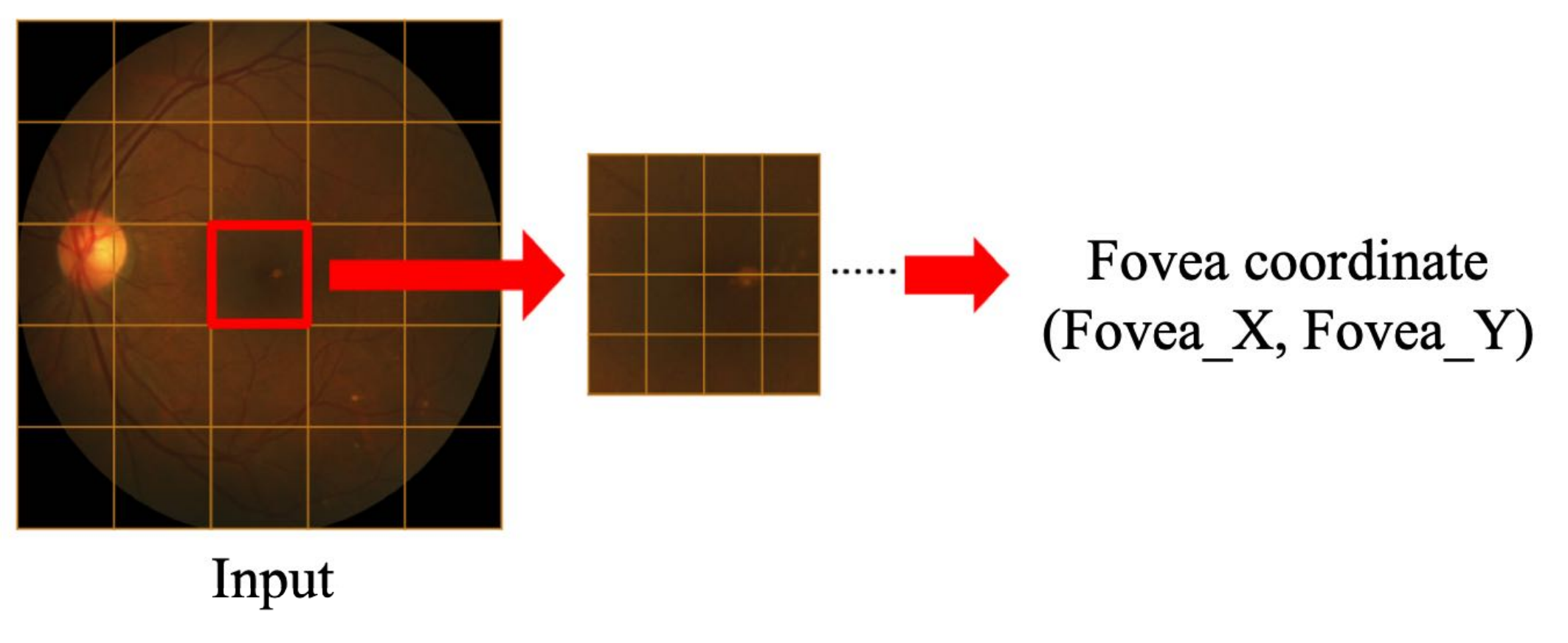}
\caption{The multiple blocks regression strategy proposed by the WWW team, which divided the image into multiple blocks, followed by making them one-hot-encoded for each block.}
\label{fig11}
\end{figure}  

The WWW team (6th) also proposed a framework based on a regression and a segmentation branches. In the regression branch, they divide the image into multiple blocks, and followed by making them one-hot-encoded for each block, as shown in Fig.~\ref{fig11}. Then, they utilized ResNet-50 to determine which block contains the fovea, and used the judged target block to make predictions in the same way, repeat three times to get the final coordinate value. 
In the segmentation branch, they took the original image size of 1024$\times$1024 and its corresponding local differential filter image as input for U-Net as two different models. And, they also trained a Mask R-CNN with the image size of 512$\times$512 to retain better spatial information. In the segmentation branch, the centroid of the segmentation was set as the prediction. Finally, for the fovea localization task, the final result is obtained by averaging the four results obtained by the above models.

\begin{figure}[t]
\centering
\includegraphics[width=1\linewidth]{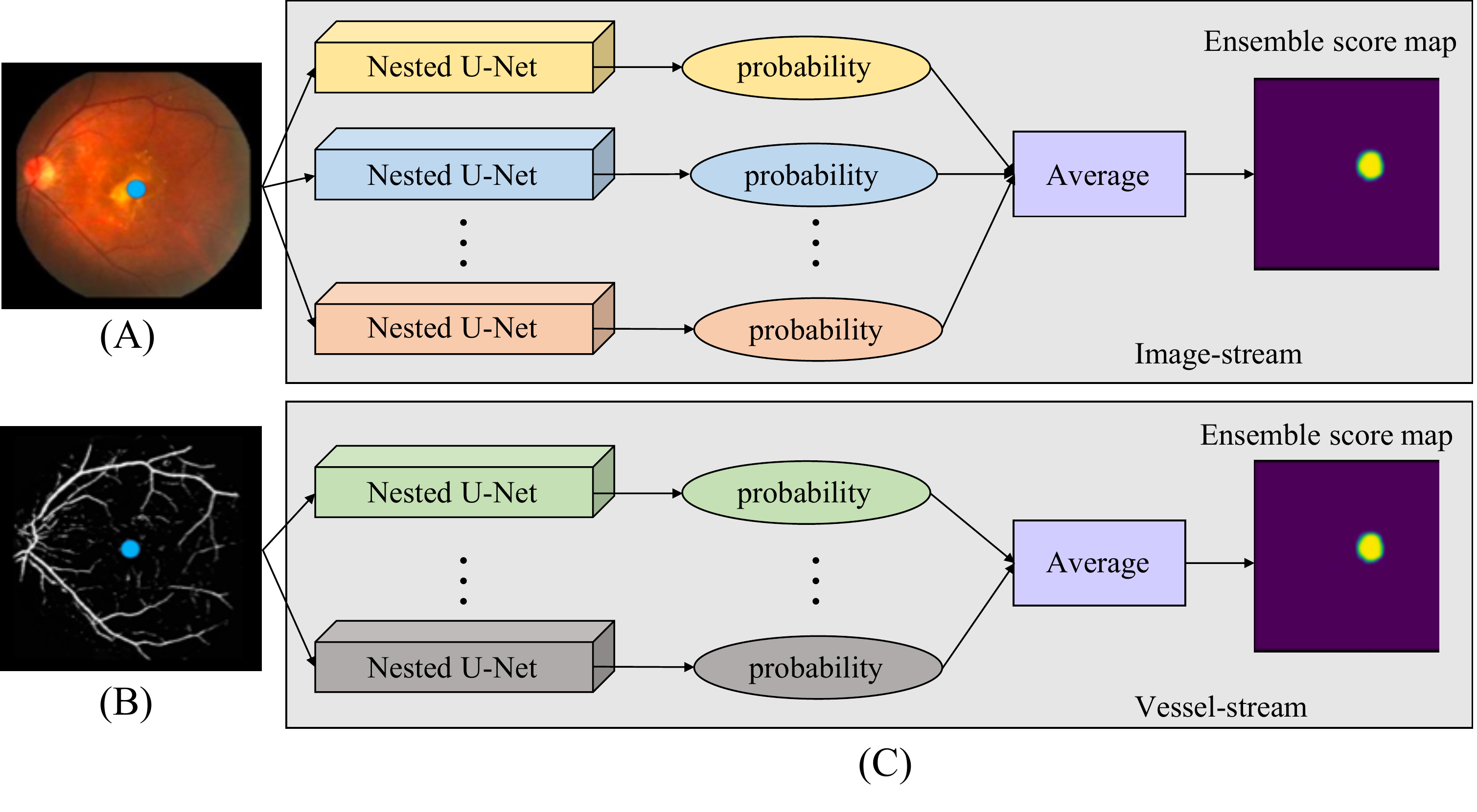}
\caption{The framework of the Voxelcloud team. (A) fundus image with a fovea point, (B) vessel segmentation mask with a fovea point, (C) the fovea segmentation framework proposed by the Voxelcloud team containing an image-stream and a vessel-stream.}
\label{fig12}
\end{figure} 

The Voxelcloud team (3rd) considered the rough fovea location could be estimated using the corresponding vessel masks. Thus, a novel fovea localization regression framework containing an image-stream structure and a vessel-steam structure was proposed, as shown in Fig.~\ref{fig12} (C). In the image-stream and the vessel-stream, they took into the fundus image and the vessel segmentation mask as input, and respectively trained six and four different models using 6-fold and 4-fold cross-validation. The architectures of the above ten models were based on Nested U-Net. Finally, the fovea regression probability maps are averaged to generate the ensemble score maps. 
During the testing phase, if the maximal value in the ensemble score map obtained by the image-stream is greater than 0.05, the fovea localization result will be calculated based on the image-stream ensemble. Otherwise, the fovea localization result will be calculated based on the vessel-stream ensemble if the maximal value in the ensemble score map obtained by the vessel-stream is greater than 0.4, if not, the fovea is thought to be not present in the image. 

In addition to the supplemented vessel information, the Airamatrix team (4th) considered the relation of the optic disc and the fovea, thus they performed the fovea segmentation task with optic disc segmentation jointly. In their framework, the encoder contains the ResNet-50 model with identity blocks while the decoder was the same as the FCN8s network. The fovea was localized accurately by obtaining the centroid coordinates of fovea segmentation masks. The ADAM-TEAM (10th) and the XxlzT team (11th) both manually marked the fovea region label according to the given coordinates, and adopted the common object detection frameworks, YOLO-v2 and Faster RCNN, to detect the target boxes covering the fovea region. Finally, they convert the output boxes to the corresponding X and Y coordinates.

The above introduction shows that two teams used the vessel segmentation information. In detail, the VUNO EYE TEAM used a GAN-based model to segment retinal vessels. Similarly, the Voxelcloud team generated the vessel masks by a U-Net with GAN Regularization. The top 3 teams and the 6th team used the ensemble method, which combined different models to achieve better results. In the training process, the VUNO EYE TEAM, Voxelcloud, and ADAM-TEAM used additional datasets, including IDRiD, REFUGE, PALM, and ARIA~\citep{chea2021classification}. Voxelcloud also used their proprietary dataset. The losses for training the regression network were the common MSE and MAE losses, and those for the segmentation network were the common BCE losses. For ADAM-TEAM and XxlzT team, the losses, such as IOU loss and Smooth L1 loss, in the YOLO-v2 and Fater RCNN frameworks were used.

\subsection{Detection and segmentation of lesions task}
\label{appendix_D}








The summary of the methods of the participating teams is shown in Table~\ref{tab5} in the main manuscript. Five teams considered the U-Net architecture or its variants. The VUNO EYE TEAM (1st), the Zasti\_AI team (2nd), the ADAM-TEAM (7th), and the TeamTiger team (8th) utilized the U-Net architecture with different encoders. Specifically, the Zasti\_AI team used Residual blocks, the TeamTiger team used EfficientNet-B0, and the ADAM-TEAM used Inception-v3, EfficientNet-B3, ResNet-50, and DenseNet-101 as the encoders, respectively. Moreover, the ADAM-TEAM used averaging method for ensembling to improve the segmentation performance. For fine segmentation of the tiny lesion, the TeamTiger team extracted the patches (256$\times$256) from each fundus image with a stride of 30 percent for training. In addition, the VUNO EYE TEAM designed both the encoder and the decoder of the U-Net, they utilized the finding network used in the AMD classification task as the encoder and adopted a decoder that consisted of depthwise separable convolutions. For each lesion segmentation task, similar to in task 1, they integrated 15 models with different parameters. Except for the traditional U-Net architecture, the Muenai\_Tim team (6th) used a nested U-Net structure~\citep{zhouUNetNestedUNet2018}, which incorporates the dense skip pathway of DenseNet. Meanwhile, they also utilized FPN, Deeplab-v3 architectures to build the feature extracting models. Finally, they ensembled these different models. 

The WWW (3rd), Airamatrix (4th), and XxlzT (10th) teams built their model based on the DeepLab-v3 architecture, which combines the advantages of spatial pyramid pooling and encoder-decoder structure for semantic segmentation task. The Airamatrix team used Xception as the backbone. The XxlzT team first trained a classification network based on ResNet50 to determine whether there was the lesion in the image, and then, for the image with lesion, they used the DeepLab-v3 framework based on ResNet101. The WWW team fused the predictions of two models (480 and 512 input sizes) to be the final segmentation result. In addition, to enhance the details for learning, the WWW team first calculated the mean image on all of the training images and followed by subtracting all images from the mean image. Second, they dilated each segmentation map by 11$\times$11 kernels to enlarge the object size for resolving the shortcoming that the small objects in the ground truth may be eliminated when down-sample the image. In this way, the small objects in the images could be successfully detected. 

The remaining CHING WEI WANG (NTUST) team (7th) utilized FCN architecture with VGG16 as encoder in  AI Explore platform to achieve the lesion segmentation. The ForbiddenFruit team (5th) adopted the FPN architecture, and they  ensembled two FPNs for each lesion type except other lesion, where the encoders were based on EfficientNet-B1 (input size 320$\times$320) and -B2 (320$\times$320) for drusen segmentation, -B2 (256$\times$256) and B1 (256$\times$256) for exudate segmentation, -B5 (384$\times$384) and -B2 (320$\times$320) for hemorrhage segmentation, -B1 (256$\times$256) and -B1 (384$\times$384) for scar segmentation. For others lesion segmentation, the encoder was based on EfficientNet-B1 (256$\times$256).

Four teams designed post-processing steps for the lesions segmentation task. The Zasti\_AI team discarded the prediction where the lesion area was less than a specific threshold found empirically. The ForbiddenFruit team took advantage of the AMD score in the post-processing step. In detail, all detections in images with an AMD score below a lesion-specific threshold $\tau$ were removed: $\tau_{scar} =0.1$, $\tau_{drusen} = \tau_{exudate} = \tau_{hemorrhage} = 10^{-3}, \tau_{other} = 10^{-4}$. The WWW and ADAM TEAM teams used the region filling and contour filling method to establish better predict results. In addition, for the training processing, only the Airamatrix team used the additional dataset DiretDBI\footnote{\url{https://www.it.lut.fi/project/imageret/diaretdb1/}} and fundus10k\footnote{\url{https://github.com/li-xirong/fundus10k}} when deal with exdute and scar lesions.

\section*{Acknowledgements}
This research was supported by the High-level Hospital Construction Project, Zhongshan Ophthalmic Center, Sun Yat-sen University (303020104), and AME Programmatic Fund (A20H4b0141). iChallenge-AMD study group includes: Hui Li (Department of Surgical Reina, Guangzhou Aier Eye Hospital, Guangzhou, China), Yingjie Li (Department of Ophthalmology, The First Hospital of Nanchang City, Nanchang, China), Renchun Xia (Department of Ophthalmology, The People's Hospital of Deyang City, Deyang, China), Chunman Yang (Department of Ophthalmology, The Second Affiliated Hospital of Guizhou Medical University, Kaili, China), Rui Zhang (Department of Ophthalmology, Qiandongnan People's Hospital, Kaili, China), Xintong Jiang (Zhongshan Ophthalmic Center, Sun Yat-sen University, Guangzhou, China), and Jian Xiong (Zhongshan Ophthalmic Center, Sun Yat-sen University, Guangzhou, China).

\bibliographystyle{unsrtnat}
\bibliography{references}

\end{document}